\documentclass[a4paper,12pt]{article}
\usepackage{jheppub} 
\usepackage[T1]{fontenc} 

\newcommand{\tx}[1]{\text {#1}}
\newcommand{\cl}[1]{\mathcal {#1}}
\newcommand{\bb}[1]{\mathbb {#1}}
\newcommand{\fk}[1]{\mathfrak {#1}}
\newcommand{\mb}[1]{\mathbf {#1}}

\def\ads{{$ AdS_{5} \times S^{5} $ }}
\def\adscp{{$ AdS_{4} \times \bb{CP}^{3} $ }}

\def\adso{{$ AdS_{3} \times S^{3} \times T^{4} $ }}

\newcommand{\Sx}{$ S $-matrix}

\DeclareMathOperator{\dd}{d\!}

\makeatletter
\newsavebox\BoxA
\newsavebox\BoxB
\newlength\LengthA

\newcommand*\obr[2][0.75]{%
	\sbox{\BoxA}{$\m@th#2$}%
	\setbox\BoxB\null
	\ht\BoxB=\ht\BoxA%
	\dp\BoxB=\dp\BoxA%
	\wd\BoxB=#1\wd\BoxA
	\sbox\BoxB{$\m@th\overline{\copy\BoxB}$}
	\setlength\LengthA{\the\wd\BoxA}
	\addtolength\LengthA{-\the\wd\BoxB}%
	\ifdim\wd\BoxB<\wd\BoxA%
	\rlap{\hskip 0.5\LengthA\usebox\BoxB}{\usebox\BoxA}%
	\else
	\hskip -0.5\LengthA\rlap{\usebox\BoxA}{\hskip 0.5\LengthA\usebox\BoxB}%
	\fi}
\makeatother

\newcommand{\bal}{\begin{equation}\begin{aligned}}
		\newcommand{\eal}{\end{aligned}\end{equation}}
\newcommand{\ov}{\over}

\newcommand{\arcsinh}{\text{arcsinh}}

\def\tp{{\widetilde p}}
\def\tE{\widetilde\cE}

\def\e{{\epsilon}}

\def\a {{\alpha}}

\def\w{{\omega}}

\def\la{\label}

\def\cE{{\cal E}}

\def\cI{{\cal I}}
\def\cJ{{\cal J}}

\def\cO{{\cal O}}

\def\bZ{{\mathbb Z}}

\renewcommand{\L}{{\scriptscriptstyle\text{L}}}
\newcommand{\R}{{\scriptscriptstyle\text{R}}}

\begin{document}
\title{\boldmath Ground state energy of twisted \adso superstring and the TBA}
\author[1]{Sergey Frolov,}
\author[2]{Anton Pribytok,}
\author[3,4,5]{Alessandro Sfondrini}

\affiliation[1]{Hamilton Mathematics Institute and School of Mathematics\\
Trinity College, Dublin 2, Ireland}
\affiliation[2]{Institut für Physik, Humboldt-Universität zu Berlin, Zum Großen Windkanal 2, 12489 Berlin, Germany}
\affiliation[3]{Dipartimento di Fisica e Astronomia, Universit\`a degli Studi di Padova, via Marzolo 8, 35131 Padova, Italy.}
\affiliation[4]{
Istituto Nazionale di Fisica Nucleare, Sezione di Padova, via Marzolo 8, 35131 Padova, Italy.}
\affiliation[5]{
Institute for Advanced Study, Einstein Drive, Princeton, New Jersey, 08540 USA.
}

\emailAdd{frolovs@maths.tcd.ie}
\emailAdd{ antons.pribitoks@physik.hu-berlin.de}
\emailAdd{alessandro.sfondrini@unipd.it}


\abstract{
\noindent
We use the lightcone \adso superstring sigma model with fermions and bosons subject to twisted boundary conditions to find the ground state energy in the semi-classical approximation where effective string tension $h$ and the light-cone momentum $L$ are sent to infinity in such a way that $\cJ\equiv L/h$ is kept fixed. 
We then analyse the ground state energy of the model by means of the mirror TBA equations for the  \adso  superstring in the pure RR background. The calculation is performed for small twist  $\mu$ with $L$ and $h$  fixed,  for large $L$ with $\mu$ and $h$ fixed, and for small $h$ with $\mu$ and $L$ fixed. 
In these limits the contribution of the gapless worldsheet modes coming from the $T^4$ bosons and fermions can be computed exactly, and is shown to be proportional to $hL/(4L^2-1)$. 
Comparison with the semi-classical result shows that the TBA equations  involve only one $Y_0$-function for massless excitations but not two as was conjectured before. Some of the results obtained are generalised to the mixed-flux  \adso superstring.
}

\maketitle

\tableofcontents

\section{Introduction and summary}

The AdS/CFT correspondence  \cite{Maldacena_1997} relates a string theory on AdS background to a conformal field theory (CFT). In particular, the energy spectrum of a string model coincides with the spectrum of scaling dimensions of the dual CFT. There are several string sigma models which are integrable on the classical and (conjecturally) quantum level, and one hopes to be able to determine their exact energy spectra.

The Thermodynamic Bethe Ansatz (TBA) approach is a powerful method to find the ground state energy (GSE) of integrable field theories  \cite{Zamolodchikov:1989cf}. It relates  the GSE of a model on a cylinder of circumference $L$ to the free energy of a so-called mirror model with temperature $T=1/L$. The mirror model is obtained by a double Wick rotation, and for non-relativistic models which appear in the AdS/CFT context it differs from the original one \cite{Arutyunov:2007tc}. The mirror TBA equations can then be used to find the spectrum of excited states via the contour deformation trick \cite{Bazhanov:1996aq,Dorey:1996re,Arutyunov:2009ax}.

The TBA approach has been successfully applied to the \ads  \cite{Arutyunov:2009zu,Arutyunov:2009ur,Bombardelli:2009ns,Gromov:2009bc} and \adscp  superstrings \cite{Bombardelli:2009xz,Gromov:2009at}, see \cite{Arutyunov:2009ga,Beisert:2010jr} for a review. The next interesting example of an integrable string sigma model is provided by \adso superstring~\cite{Cagnazzo:2012se}. The superstring background can be supported by both Ramond-Ramond (RR) and Neveu- Schwarz-Neveu-Schwarz (NSNS) fluxes \cite{Larsen:1999uk,OhlssonSax:2018hgc}. 
The properties of the general model are poorly understood, and its dual CFT is unknown. 
 In the case of pure NSNS background the superstring theory in the conformal gauge is a level-$k$ supersymmetric Wess-Zumino-Novikov-Witten (WZNW) model, and its spectrum  can be found by using CFT methods \cite{Maldacena:2000hw}. The pure NSNS superstring is believed to be dual to symmetric product orbifold CFTs \cite{Giribet:2018ada,Eberhardt:2018ouy,Eberhardt:2021vsx}.

The pure RR \adso Green-Schwarz (GS)  superstring can be analysed in the uniform lightcone gauge by using the methods developed for the \ads superstring~\cite{Arutyunov:2009ga}. Since in this gauge the density of the momentum $p^+$ is set to 1, the total lightcone momentum $P^+$ is equal to circumference $L$ of the cylinder where the gauged-fixed model lives. Taking the decompactification limit $L\to\infty$, one gets a model on a plane. The symmetry algebra $ \fk{psu}(1,1|2)_{\L} \oplus \fk{psu}(1,1|2)_\R $ of the \adso model is broken  in the lightcone gauge to  
\begin{equation}
\label{eq:lcsymm}
    (\fk{psu}(1|1)_\L \oplus \fk{psu}(1|1)_\R)_{\rm c.e} \oplus (\fk{psu}(1|1)_\L \oplus \fk{psu}(1|1)_\R)_{\rm c.e}\,,
\end{equation}
where ``c.e.'' indicates a central extension. Like in $AdS_5\times S^5$, two central charges vanish for states satisfying the level-matching condition, see \textit{e.g.}~\cite{Sfondrini:2014via} for details. There are two more symmetries that play a distinguished  role in the integrability constructions, which we denote as $\fk{su}(2)_\bullet\oplus\fk{su}(2)_\circ\cong\fk{so}(4)$ and correspond to the local $\fk{so}(4)$ symmetry of $T^4$  under which four bosons and all spinors of the Green-Schwarz string transform. Specifically, $\fk{su}(2)_\bullet$ acts as an automorphism on~\eqref{eq:lcsymm} while $\fk{su}(2)_\circ$ commutes with~\eqref{eq:lcsymm} (and indeed with $\fk{psu}(1,1|2)_\L\oplus\fk{psu}(1,1|2)_\R$).

Asymptotic particles transform in four-dimensional short representations of the lightcone algebra which differ by the value $M\in\bZ$ of an external $U(1)$ automorphism \cite{Borsato:2013hoa}. In the large string tension expansion the absolute value of $M$ is identified with the mass of a particle. Particles with $M=2,3,\dots$ and $M=-2,-3,\dots$ are bound states of particles with $M=+1$ and  $M=-1$, respectively  \cite{Borsato:2013hoa}, and the dispersion relation is
\begin{equation}
    E(p)=\sqrt{M^2+4h^2\sin^2(p/2)}\,,
\end{equation}
where the coupling~$h$ is related to the string tension.
 The worldsheet S-matrix of  the lightcone model is fixed by symmetries up to several ``dressing'' factors which obey crossing equations \cite{Borsato:2014hja}. 
 A solution to the crossing equations was recently proposed in \cite{Frolov:2021fmj},  and used to analyse the properties of the mirror \adso model, as well as to derive the mirror TBA equations \cite{Frolov:2021bwp}.    The Bethe-Yang equations of both models involve massive momentum-carrying excitations corresponding to $|M|\ge 1$, two types of gapless ($M=0$) momentum-carrying excitations, and two types of auxiliary excitations which account for the $(\fk{psu}(1|1)_\L \oplus \fk{psu}(1|1)_\R)_{\rm c.e} \oplus (\fk{psu}(1|1)_\L \oplus \fk{psu}(1|1)_\R)_{\rm c.e}$ structure of the multiplets~\cite{Borsato:2012ss,Seibold:2022mgg}. More specifically auxiliary excitations are related to~$\fk{su}(2)_\bullet$, while the fact that we have two multiplets with~$M=0$ is related to $\fk{su}(2)_\circ$.
 
 In the  TBA approach it is necessary to introduce particle/hole distributions for all types of excitations that appear in the thermodynamic limit; such distributions are encoded in the so-called Y-functions.
 Moreover, we consider a mirror model with dispersion
 \begin{equation}
     \tilde{E}(\tilde{p})=2\text{arcsinh}\left(\frac{\sqrt{M^2+\tilde{p}^2}}{2h}\right)\,.
 \end{equation}
The mirror TBA equations are written as functions of the mirror momentum $\tilde{p}$ (or a suitable rapidity) and feature the following $Y$-functions~\cite{Frolov:2021bwp}:
i) $Y_Q$  of  $Q$-particles with $M=Q\ge1$, ii) $\overline{Y}_Q$  of  $\overline{Q}$-particles with $M=-\overline{Q}\le-1$, iii) $Y_0^{(\dot\a)}$, $\dot\a = \dot{1}, \dot{2} $ of massless particles,  iv) $Y_\pm^{(\a)}$, $\a = 1,2$ of auxiliary excitations. 
The {\adso} ground state energy receives contributions from both massless and  massive excitations
\begin{equation}\label{GSE_Complete}
	\begin{aligned}
		E(h, L,\mu) =&  -\sum_{\dot\a = 1}^{N_0}\,\int\displaylimits_{-\infty}^{\infty} \frac{d \widetilde{p}}{2 \pi} \log \left(1+Y_0^{(\dot\a)}\right)\\
  &\qquad -\sum_{Q = 1}^{+\infty} \, \int\displaylimits_{-\infty}^{\infty} \frac{d \widetilde{p}}{2 \pi}\log \left[ \left(1+Y_Q\right)\left(1+\overline{Y}_Q\right) \right] 	\,,
  \end{aligned}
\end{equation}
where  $h, L,\mu$ are the three parameters of the TBA equations: i) effective string tension $h$, ii) light-cone momentum $L$, iii) twist  $\mu$. In the temporal gauge $L$ is identified with the charge $ J $ corresponding to a $ U(1)$ isometry of $ S^{3} $ and acquires only integer or half-integer values.
The twist $\mu$ allows one to consider the lightcone string sigma model with twisted boundary conditions on the fields that are charged under $\fk{su}(2)_\bullet$ or $\fk{su}(2)_\circ$. In the untwisted string model, bosons and fermions have periodic boundary conditions (in absence of winding) so that the GSE vanishes, as expected from supersymmetry.
$N_0$ in \eqref{GSE_Complete} denotes the number of $Y_0^{(\dot\a)}$-functions for gapless momentum-carrying roots which, as has been mentioned above, was chosen to be equal to 2 in  \cite{Frolov:2021bwp}. However, the  $Y_0^{(\dot\a)}$-functions are the same for any $\dot\a$, and it leaves a possibility that only one $Y_0$-function should appear in the TBA equations. To address this question in this work we analyse the ground state energy of the  \adso superstring in two different ways.  

We begin with the lightcone sigma model  with fermions and bosons subject to twisted boundary conditions, and calculate the GSE 
 in the semi-classical approximation where effective string tension $h$ and the light-cone momentum $L$ are sent to infinity in such a way that $\cJ\equiv L/h$ is kept fixed. The GSE is given by the sum of the contributions of the massless and massive particles, $E = E_0+E_{\rm m}$. 
  
  The massless particles contribution to the GSE is found to be  given by
\bal\label{Emasslessintro}
E_0=-\frac{\mu ^2}{\pi\cJ }+\frac{|\mu+\mu'| + |\mu-\mu'| -2|\mu'| }{\cJ }\,,
\eal
where $\mu$ is the twist of massive fermions and massless bosons due to the $su(2)_\bullet$ symmetry of the model while $\mu'$ is used to 
twist massless fermions and bosons thanks to the $su(2)_\circ$ symmetry. The twists take values between $-\pi$ and $\pi$. If we set  $\mu'=0$ we get a term linear in $|\mu|$ in \eqref{Emasslessintro}. It is unclear how such a term can be obtained from the TBA. On the other hand if  $|\mu|\le|\mu'|$ then
\bal\label{Emasslessintro2}
E_0
=-\frac{\mu ^2}{\pi\cJ }\quad \text{if}\quad |\mu|\le|\mu'|\,,
\eal
and for small $\mu$ the $\mu^2$ dependence is the expected one from the TBA analysis  performed for the \ads superstring  in  \cite{Frolov:2009in}. For finite $\mu$, however, one would expect from the TBA that $\mu^2$ would be replaced by $4\sin^2(\tfrac{\mu}{2})$.

The massive particles contribution to the GSE can be found in the large $\cJ$ limit where we get
\bal\label{Emassiveintro}
E_{\rm m}&=-4\sin^2(\tfrac{\mu }{2})\,\sqrt{\frac{2}{\pi }}\,\frac{ e^{-\cJ} }{\sqrt{\cJ}} + \cO\left(e^{-J}/\cJ^{3/2}\right)\,,
\eal
and it is of the expected form.

We continue our analysis by solving the mirror TBA equations in three regimes where $Y$-functions of massive and massless particles are small. 
First we consider  the small-twist regime ($\mu\ll1$) around a BPS vacuum with  $L$ and $h$ fixed. Then, we analyse the regime of large~$L$ with $\mu$ and $h$ fixed. Finally, the regime of small~$h$ with $\mu$ and $L$ fixed is discussed. The analysis follows closely the one performed for the \ads superstring  in \cite{Frolov:2009in}. There are two main differences in comparison to the \ads case. First, the GSE is finite for $L>1$ for \adso and for $L>2$ for \ads. Second, in the \ads case there are no massless modes while in the \adso case the massless worldsheet modes come from the $T^4$ bosons and fermions, and their contribution in these limits can be computed exactly. The massless mode contribution to the GSE \eqref{GSE_Complete} appears to be  equal to 
\bal\label{GSE_massless}
E_0^{\makebox{\tiny TBA}}(h,L,\mu) \approx - \frac{4}{\pi} \sin ^2(\tfrac{\mu }{2})\, \frac{N_0 h L}{L^2-{1\ov 4}} \,,
\eal
where for small $\mu$ one clearly replaces $4\sin^2\frac{\mu}{2}\to \mu^2$. The contribution appears at linear order in $h$, and for large $L$ it behaves as $1/L$. This is in complete agreement with the recent results obtained in \cite{Brollo:2023pkl} where the string spectrum was analysed in the tensionless limit $h\to0$ of the \adso TBA equations. For small $\mu$ the solution of the TBA equations is also valid in the 
limit  $h\to\infty$, $L\to\infty$ and $\cJ\equiv L/h$ fixed where we find
\bal\label{GSEmasslessintro}
E_0^{\makebox{\tiny TBA}}(\cJ,\mu) = - N_0\frac{\mu ^2}{\pi\cJ }\,.
\eal
Comparing \eqref{GSEmasslessintro} with \eqref{GSE_massless}, we conclude that to have the agreement with the semiclassical string model calculation one has to set $N_0$ to 1. This is the main result of the paper.

The paper is structured as follows. In section \ref{lcstring} we calculate the GSE 
 in the semi-classical approximation, and derive 
\eqref{Emasslessintro} and \eqref{Emasslessintro2}. 
 In section  \ref{Sec:Small_mu} we  look into perturbative Ansätze for $ Y $-functions in the small twist $ \mu $  regime around BPS vacuum, and obtain \eqref{GSE_massless} and  show that the contribution of the massive modes is given by 
\bal\label{GSE_mass}
E_{\rm m}^{\makebox{\tiny TBA}}(h,L,\mu) \approx - \frac{\mu^2}{\pi} \cI(h,L) \,,
\eal
where $ \cI(h,L)$ is the following sum\footnote{Note that in the \ads case the GSE up to a numerical factor is given by \eqref{GSE_mass} but the summand contains an extra factor of $Q^2$: 
$\cI_{AdS_5}(h,L)\equiv  \sum_{Q = 1}^{+\infty} Q^2 \int_{-\infty}^{\infty} \mathrm{d}  \widetilde{p}\, e^{-L \tilde{\cl{E}}_{Q}}\,.
$
}
\bal
\cI(h,L)\equiv  \sum_{Q = 1}^{+\infty} \int\displaylimits_{-\infty}^{\infty} \mathrm{d}  \widetilde{p}\,  e^{-L \tilde{\cl{E}}_{Q}}\,,\quad \widetilde{\cl{E}}_{Q} = 2 \, \arcsinh\left( \dfrac{\sqrt{ \widetilde{p}^{2} + Q^{2}}}{2 h} \right)\,.
\eal
The sum cannot be evaluated in a closed form but one can show that it is convergent for $L>1$, and it can be expanded in a power series in $h$ with explicit expansion coefficients, see \eqref{S_Massive}.
The series begins with a term of order $h^{2L}$, and therefore for small $h$ the main contribution comes from massless modes and is given by \eqref{GSE_massless}.

The sum can also be computed for $h\gg L$ with $L$ kept fixed where we get
\bal\label{hggL}
\cI(h\gg L,L)&\approx h^2{\pi\ov L^2-1}\,,\quad  
\cI_{AdS_5}(h\gg L,L)\approx h^4\frac{3 \pi }{L^4-5 L^2+4} \,,
\eal
and in addition  we have to assume that $\mu\, h^L\ll 1$ otherwise $Y$-functions will be large. 
In this case massive modes provide the leading contribution, and there is a substantial difference between the \adso and \ads superstrings.

Finally, for small $\mu$ we can consider the limit $L\to\infty$, $h\to\infty$ while $\cJ\equiv L/h$ is kept fixed. Then we get
\begin{equation}
\begin{aligned}
\cI(\cJ )&
= \int\displaylimits_{-\infty}^{\infty} \mathrm{d} p\,  {1\ov 4 \sinh ^2\left(\frac{1}{2} \cJ \sqrt{p^2+1}\right)}\,,\\
\cI_{AdS_5}(\cJ ) &=  \int\displaylimits_{-\infty}^{+\infty} \dd p \, \dfrac{2 + \cosh \left(\cJ \sqrt{p^2+1}\right)}{8 \sinh^{4}\left(\frac{\cJ}{2} \sqrt{p^2+1} \right)}\,. 
\end{aligned}
\end{equation}
For finite $\cJ$ the integral cannot be computed analytically but for large and small $\cJ$ one finds
\bal
\cI(\cJ\gg1 )&\approx 
\frac{\sqrt{2\pi} e^{-\cJ}}{ \sqrt{\cJ }} \,,\qquad \cI_{AdS_5}(\cJ\gg1 )\approx \frac{\sqrt{2\pi} e^{-\cJ}}{ \sqrt{\cJ }} \,,
\eal
\bal
\cI(\cJ\ll1 )&\approx {\pi\ov \cJ^2}\,,\qquad \cI_{AdS_5}(\cJ\ll1 )\approx  {3\pi\ov \cJ^4}\,.
\eal
As expected, the contribution of massive particles is exponentially suppressed for large $\cJ$ while massless particles provide the leading contribution of order $1/\cJ$. On the other hand for small $\cJ$ massive particles contribute more to the GSE.

In section \ref{Sec:LCA} we find the leading order solution of the TBA system for large $L$ with $\mu$ and $h$ fixed. The same solution also applies to the case of small $h$ with $\mu$ and $L$ fixed.  We show that the GSE is still given by the sum of \eqref{GSE_massless} and \eqref{GSE_mass} with $\mu^2$ replaced by $4\sin^2\mu/2$. We also check that the same expression for the GSE is obtained from the generalised Lu\"scher formula 
\cite{Luscher:1985dn,Janik:2007wt,Bajnok:2008bm,Hatsuda:2008na,Ahn:2011xq,Bombardelli:2013yka}.
 For large $L$ with $h$ fixed the main contribution of massive particles comes from $Q=1$, and 
 we get
\bal
\cI(h,L\gg h)&\approx \frac{\sqrt{\pi } \sqrt[4]{4 h^2+1} \left(\sqrt{\frac{1}{4
			h^2}+1}+\frac{1}{2 h}\right)^{-2 L}}{\sqrt{L}} \,.
\eal
We see that for any $h$ the massive contribution is exponentially suppressed while the massless one is leading and proportional to $h/L$. 
For small and large $h$ the formula simplifies
\bal
\cI(h\ll1,L\gg h)&\approx \sqrt{\frac{\pi} {L}}\, h^{2 L} \,,\quad
\cI(h\gg1,L\gg h)\approx \sqrt{2\pi \frac{h}{L}}\, e^{- {L\ov h}} \,.
\eal
These formulae with $\mu=\pi$ provide us with the GSE of  the odd-winding number sector with anti-periodic fermions and nonsupersymmetric vacuum.

In section \ref{mixedflux} we  generalise the semiclassical consideration in section \ref{lcstring} to the twisted lightcone string on one-parameter family of mixed-flux $AdS_3\times S^3\times S^3\times S^1$  backgrounds. The corresponding  lightcone string sigma model was considered in \cite{Dei:2018yth}, and it  contains the  mixed-flux  \adso superstring when one sphere blows up. We derive the general formula \eqref{Eads3s3s3s1c} for the GSE, and specialise it to the \adso case in \eqref{Eads3s3t4d}.
 Then, assuming that the TBA equations for the mixed-flux  \adso superstring are similar to the pure RR ones, we reproduce the semiclassical result from $Y_Q$-functions.

We make concluding remarks in section \ref{conclusion}. The full set of the \adso TBA equations is collected in appendix \ref{Apx:TBASystem}, and necessary kernels are defined in appendix \ref{Apx:S_K_Relations}.

\section{Lightcone string GSE}\label{lcstring}

The ground state energy of the twisted {\adso} lightcone superstring can be  analysed by  studying the perturbative expansion of the lightcone action in terms of inverse tension
\begin{equation}\label{action}
S = \int \dd \tau\int_0^\cJ \dd \sigma \, \cl{L}\,,\qquad	\cl{L} = \cl{L}_{2} + \dfrac{1}{h}\cl{L}_{4} + \dfrac{1}{h^{2}}\cl{L}_{6} + \dots\,,
\end{equation}
where $\cJ\equiv {L\ov h}$, and the orders with odd field configuration are absent due to perturbative properties of $ AdS_{n} \times S^{n} $ spaces \cite{Sundin:2014ema}. The quadratic Lagrangian takes the form 
\begin{equation}\label{lagr2}
	\cl{L}_{2} = \cl{L}_{0} + \cl{L}_{\tx{m}}\,,
\end{equation}
with terms describing massless and massive sectors correspondingly 
\begin{equation}\label{lagr0}
	\begin{aligned}
		\cl{L}_{0} & = \left|\partial_i u^{\dot\a}\right|^2 + i \bar{\chi}_L^{\dot\a} \partial_{-} \chi_L^{\dot\a}+i \bar{\chi}_R^{\dot\a} \partial_{+} \chi_R^{\dot\a} \,, \\[1ex]
		\cl{L}_{\tx{m}} & = \left|\partial_i {x^a}\right|^2-|{x^a}|^2 + i \bar{\chi}_L^\alpha \partial_{-} \chi_L^\alpha + i\bar{\chi}_R^\alpha \partial_{+} \chi_R^\alpha -  \bar{\chi}_L^\alpha \chi_R^\alpha - \bar{\chi}_R^\alpha \chi_L^\alpha\,,
	\end{aligned}
\end{equation}
where $ u^{\{ \dot 1,\dot 2 \}} $ and $ \chi^{\{ \dot 1,\dot 2 \}} $ are massless bosons and fermions, whereas $ x^{\{ 1,2 \}} $ and $ \chi^{\{ 1,2 \}} $ are the massive ones. 

In terms of these fields the action \eqref{action} is invariant under the U(1) transformations corresponding to SU(2)$_\bullet$
\begin{equation}
    \chi^{\alpha}\to e^{i\mu}\chi^{\alpha},\qquad
    \chi^{\dot{\alpha}}\to \chi^{\dot{\alpha}},\qquad
    u^{\dot\a}\to e^{i\mu}u^{\dot \a},\qquad
    x^a\to x^a\,,
\end{equation}
and, under the U(1) transformations corrsponding to SU(2)$_\circ$,
\begin{equation}
    \chi^{\alpha}\to \chi^{\alpha},\qquad
    \chi^{\dot \alpha}\to e^{\pm i\mu'}\chi^{\dot \alpha},\qquad
    u^{\dot\alpha}\to e^{\pm i\mu'}u^{\dot \alpha},\qquad
    x^a\to x^a\,,
\end{equation}
where the sign of the twist is positive for $\dot{\alpha}=\dot{1}$ and negative for $\dot{\alpha}=\dot{2}$.
The invariance can be used to impose twisted boundary conditions on the fields.

\paragraph{Massless.} We first analyse the massless part $ \cl{L}_{0} $ \eqref{lagr0}. For generality we assume that both the massless bosonic and fermionic fields satisfy twisted boundary conditions which we choose to be
\bal\label{twistcondmassless}
	u^{\dot 1}(\sigma + \cJ,\tau) &= e^{ i (\mu +\mu')}  u^{\dot 1}(\sigma,\tau)\,,\qquad &u^{\dot 2}(\sigma + \cJ,\tau) &= e^{ i (\mu -\mu')} u^{\dot 2}(\sigma,\tau)\,,
\\
\chi^{\dot 1}(\sigma + \cJ,\tau) &= e^{+ i \mu'} \chi^{\dot 1}(\sigma,\tau)\,,\qquad &\chi^{\dot 2}(\sigma + \cJ,\tau) &= e^{- i \mu'} \chi^{\dot 2}(\sigma,\tau)\,,
\eal
where $\mu$ and $\mu'$ are the twists of SU(2)$_\bullet$ and SU(2)$_\circ$, respectively and $-\pi\le\mu\,,\,\mu'\le \pi$.

The twisted boundary conditions imply the following mode expansion of the bosonic fields 
\bal\label{modesbosmassless}
u^{\dot 1}(\sigma,\tau) &= {1\ov\sqrt\cJ} \sum_{n=-\infty}^\infty u^{\dot 1}_{n} \, e^{2 \pi i (n+\hat\mu+\hat\mu') \frac{\sigma}{\cl{J}}}\,, \quad u^{\dot 2}(\sigma,\tau) = {1\ov\sqrt\cJ} \sum_{n=-\infty}^\infty u^{\dot 2}_{n} \, e^{2 \pi i (n+\hat\mu-\hat\mu') \frac{\sigma}{\cl{J}}}\,,
\eal
where $\hat\mu = {\mu\ov 2\pi}$, $\hat\mu' = {\mu'\ov 2\pi}$.

 For massless fermions we assume for definiteness that $0\le\mu'\le\pi$, and choose the following mode expansion
\bal\label{masslessmodes}
\chi^{\dot 1}_L(\sigma,\tau) &= {1\ov\sqrt\cJ} \left(\sum_{n=0}^\infty c^{{\dot 1}\dagger}_{nL} \, e^{2 \pi i (n+\hat\mu') \frac{\sigma}{\cl{J}}}+\sum_{n=1}^\infty d^{{\dot 1}}_{nL} \, e^{-2 \pi i (n-\hat\mu') \frac{\sigma}{\cl{J}}}\right)\,,
\\
\chi^{\dot 2}_L(\sigma,\tau) &= {1\ov\sqrt\cJ} \left(\sum_{n=1}^\infty c^{{\dot 2}\dagger}_{nL} \, e^{2 \pi i (n-\hat\mu') \frac{\sigma}{\cl{J}}}+\sum_{n=0}^\infty d^{{\dot 2}}_{nL} \, e^{-2 \pi i (n+\hat\mu') \frac{\sigma}{\cl{J}}}\right)\,,
\\
\chi^{\dot 2}_R(\sigma,\tau) &= {1\ov\sqrt\cJ} \left(\sum_{n=0}^\infty c^{{\dot 2}\dagger}_{nR} \, e^{-2 \pi i (n+\hat\mu') \frac{\sigma}{\cl{J}}}+\sum_{n=1}^\infty d^{{\dot 2}}_{nR} \, e^{2 \pi i (n-\hat\mu') \frac{\sigma}{\cl{J}}}\right)\,,
\\
\chi^{\dot 1}_R(\sigma,\tau) &= {1\ov\sqrt\cJ} \left(\sum_{n=1}^\infty c^{{\dot 1}\dagger}_{nR} \, e^{-2 \pi i (n-\hat\mu') \frac{\sigma}{\cl{J}}}+\sum_{n=0}^\infty d^{{\dot 1}}_{nR} \, e^{2 \pi i (n+\hat\mu') \frac{\sigma}{\cl{J}}}\right)\,,
\eal
Substituting the mode expansions of the bosons and fermions into the Lagrangian, we get
\begin{equation}\label{L0modes}
\begin{aligned}
 \cl{L}_{0} &= \sum_{n=-\infty}^\infty \left(\dot{\bar{{u}}}_{n}^{\dot\a} \, \dot{{u}}_{n}^{\dot\a}
- \left( {2\pi (n+\hat\mu+\hat\mu')\ov \cJ}\right)^2 \, \bar{{u}}_{n}^{\dot1} {u}_{n}^{\dot1} - \left( {2\pi (n+\hat\mu-\hat\mu')\ov \cJ}\right)^2 \, \bar{{u}}_{n}^{\dot2} {u}_{n}^{\dot2}\right)\\[1ex]
& + \sum_{n=0}^\infty \left(i \, c^{{\dot\a}\dagger}_{nL}\dot c^{{\dot\a}}_{nL}+ i \,c^{{\dot\a}\dagger}_{nR}\dot c^{{\dot\a}}_{nR} \, +
{2\pi(n+\hat\mu')\ov \cJ}(c^{{\dot1}}_{nL}c^{{\dot1}\dagger}_{nL}+c^{{\dot2}}_{nR}c^{{\dot2}\dagger}_{nR}) \right.
\\
&\hspace{4.7cm}\left.+\,{2\pi(n+1-\hat\mu')\ov \cJ}(c^{{\dot2}}_{nL}c^{{\dot2}\dagger}_{nL}+c^{{\dot1}}_{nR}c^{{\dot1}\dagger}_{nR})\right)
 \\
&+\sum_{n=1}^\infty \left( i \, d^{{\dot\a}\dagger}_{nL}\dot d^{{\dot\a}}_{nL}+ i \, d^{{\dot\a}\dagger}_{nR}\dot d^{{\dot\a}}_{nR}-{2\pi(n-\hat\mu')\ov \cJ}(d^{{\dot1}\dagger}_{nL}d^{{\dot1}}_{nL}+d^{{\dot2}\dagger}_{nR}d^{{\dot2}}_{nR}) \right.
\\ 
&\hspace{4.7cm}\left. -\,{2\pi(n-1+\hat\mu')\ov \cJ}(d^{{\dot2}\dagger}_{nL}d^{{\dot2}}_{nL} +d^{{\dot1}\dagger}_{nR}d^{{\dot1}}_{nR})\right)\,.	 
\end{aligned}
\end{equation}
Choosing $c,d$ and $c^\dagger,d^\dagger$ as annihilation and creation operators, respectively, we find 
the frequencies which contribute to the GSE
\bal
\w_n^{{\dot1}b}&= {2\pi\ov \cJ} |n+\hat\mu+\hat\mu'|\,,\quad \w_n^{{\dot2}b}= {2\pi\ov \cJ} |n+\hat\mu-\hat\mu'|\,,\quad n\in \bZ\,,
\\
\w_n^{{\dot1}f}&= {2\pi\ov \cJ} (n+\hat\mu')\,,\quad \w_n^{{\dot2}f}= {2\pi\ov \cJ} (n+1-\hat\mu')\,,\quad n=0,1,\ldots, \infty\,.
\eal
Thus, the GSE is given by
\bal\label{Emassless}
E_0&= \sum_{n=-\infty}^\infty\left(\w_n^{{\dot1}b} +\w_n^{{\dot2}b} \right)  - 2\sum_{n=0}^\infty \left(\w_n^{{\dot1}f} +\w_n^{{\dot2}f} \right)
\\
&={2\pi \ov \cJ}\left( |\hat\mu+\hat\mu'|+\sum_{n=0}^\infty  (n+1+\hat\mu+\hat\mu')\right)+{2\pi \ov \cJ}\sum_{n=0}^\infty  (n+1-\hat\mu-\hat\mu')
\\
&+{2\pi \ov \cJ}\left( |\hat\mu-\hat\mu'|+\sum_{n=0}^\infty  (n+1+\hat\mu-\hat\mu')\right)+{2\pi \ov \cJ}\sum_{n=0}^\infty  (n+1-\hat\mu+\hat\mu')
\\
& - {4\pi \ov \cJ}\sum_{n=0}^\infty (n+\hat\mu') - {4\pi \ov \cJ}\sum_{n=0}^\infty (n+1-\hat\mu')
\,.
\eal
To compute the divergent series we can use the Hurwitz zeta-function
\bal
\zeta(s,a)=\sum_{n=0}^\infty {1\ov (n+a)^s}\,,
\eal
and get\footnote{The same answer is obtained by using a more straightforward regularisation
\[
E_0=\sum_{n=-\infty}^\infty\left(\w_n^{{\dot1}b}e^{-\epsilon\, \w_n^{{\dot1}b}} +\w_n^{{\dot2}b} e^{-\epsilon\, \w_n^{{\dot2}b}}\right)  - 2\sum_{n=0}^\infty \left(\w_n^{{\dot1}f}e^{-\epsilon\, \w_n^{{\dot1}f}} +\w_n^{{\dot2}f}e^{-\epsilon\, \w_n^{{\dot2}f}} \right)\,.
\]
}
\begin{equation}
\begin{aligned}
\label{Emassless2a}
E_0&={2\pi \ov \cJ}\Big[ |\hat\mu+\hat\mu'| + |\hat\mu-\hat\mu'|
\\
&\qquad\qquad+ \zeta(-1, 1+\hat\mu+\hat\mu')+ \zeta(-1, 1-\hat\mu-\hat\mu')\\
&\qquad\qquad+ \zeta(-1, 1+\hat\mu-\hat\mu')+ \zeta(-1, 1-\hat\mu+\hat\mu')
\\
&\qquad\qquad-2\zeta(-1, \hat\mu')-2\zeta(-1, 1-\hat\mu')\Big] \\
&={2\pi \ov \cJ}( |\hat\mu+\hat\mu'| + |\hat\mu-\hat\mu'|-2\hat\mu^2 -2\hat\mu' )\,.
\end{aligned}
\end{equation}
Since the GSE has to be symmetric under $\mu'\to -\mu'$, we finally find
\bal\label{Emassless2}
E_0&=-\frac{\mu ^2}{\pi\cJ }+\frac{|\mu+\mu'| + |\mu-\mu'| -2|\mu'| }{\cJ }\,.
\eal
We see that for $| \mu|\ge|\mu'|$ we get a term linear in $|\mu|-|\mu'|$ while for $| \mu|\le|\mu'|$
the dependence on $\mu'$ and the linear term in $|\mu|$ disappears, and we obtain
\bal\label{Emassless3}
E_0
=-\frac{\mu ^2}{\pi\cJ }\quad \text{if}\quad |\mu|\le|\mu'|\,.
\eal
We expect that when all fields are periodic $\mu=0$ and $E_0=0$. This suggests that we should identify the twist in the TBA equations with  $\mu'=\mu$. 

\vspace{0.5cm}
\paragraph{Massive.} The massive bosonic fields are periodic while the fermionic fields satisfy the twisted boundary conditions 
\begin{equation}\label{twistcondmass}
	x^a(\sigma + \cJ,\tau) = x^a(\sigma,\tau)\,,\qquad 	\chi^\alpha(\sigma + \cJ,\tau) = e^{ \pm i \mu} \chi^\alpha(\sigma,\tau)\,,\quad -\pi\le\mu\le \pi\,,
\end{equation}
where the sign in the exponent depends on the type of a fermion.

The twisted boundary conditions imply the following mode expansion of the fields 
\bal\label{modes}
	x^a(\sigma,\tau) &=\frac{1}{\sqrt\cJ} \sum_{k} {x}^a_{k} \, e^{2 \pi i k \frac{\sigma}{\cl{J}}}\,,
	\qquad
	\chi^\alpha(\sigma,\tau) &= {1\ov\sqrt\cJ} \sum_{n} \chi_{n}^\alpha \, e^{2 \pi i (n+\hat\mu) \frac{\sigma}{\cl{J}}}\,.
	\eal
The calculation of the contribution to the GSE from the massive bosons and fermions  follows the standard lines, and is given by
\bal\label{Emassive}
E_m&= 2\sum_{k=-\infty}^\infty \omega_k^b  - 2\sum_{k=-\infty}^\infty \omega_{k}^f 
\,,
\eal
where 
\bal
 \omega_k^b= \sqrt{\left(\frac{2 \pi  k}{\cl{J}}\right)^2 + 1}\,,\quad  \omega_{k}^f= \sqrt{\left(\frac{2 \pi (k+\hat\mu)}{\cl{J}}\right)^2 + 1}\,.
\eal
We want to evaluate \eqref{Emassive} for $\cJ\gg1$.\footnote{Note that for finite $\cJ$ one also has to take into account the contribution of bound states to the GSE.} We use the following regularisation
\bal\label{Emassive2}
E_m&= 2\sum_{k=-\infty}^\infty \omega_k^b e^{-\epsilon\,\omega_k^b}  - 2\sum_{k=-\infty}^\infty \omega_{k}^fe^{-\epsilon\,\omega_{k}^f} 
=2\sum_{k=-\infty}^\infty \left(\omega_k^b e^{-\epsilon\,\omega_k^b}  -  \omega_{k}^fe^{-\epsilon\,\omega_{k}^f}\right)
\,.
\eal
By applying the Poisson summation formula, we obtain
\bal\label{Emassivemu}
E_m&= 2\sum_{w=-\infty}^\infty \left(F_w^b   - F_w^f\right)=2F_0^b   - 2F_0^f+ 2\sum_{w\neq0} \left(F_w^b   - F_w^f\right)
\,,
\eal
where
\bal
F_w^b &= \int_{-\infty}^\infty dk\, \omega_k^b e^{-\epsilon\,\omega_k^b} e^{-i2\pi wk} =\frac{\cl{J}}{2 \pi}\int_{-\infty}^\infty dx\, \sqrt{x^2 + 1}e^{-\epsilon\, \sqrt{x^2 + 1}} e^{-i \cJ wx}\,,
\eal
and
\bal
F_w^f &= \int_{-\infty}^\infty dk\, \omega_k^f e^{-\epsilon\,\omega_k^f} e^{-i2\pi wk} \\
&=e^{i w\mu}\frac{\cl{J}}{2 \pi}\int_{-\infty}^\infty dx\, \sqrt{x^2 + 1}e^{-\epsilon\, \sqrt{x^2 + 1}} e^{-i \cJ wx} =e^{i w\mu}\,F_w^b\,.
\eal
Thus, $F_0^b  =F_0^f$ and we only need to calculate $F_w^b$ for $w\neq 0$. Integrating by parts twice, we find
\bal
F_w^b 
&=-\frac{1}{2 \pi  \cJ w^2}\int_{-\infty}^\infty dx\,  \frac{e^{-i \cJ w x}}{ \left(x^2+1\right)^{3/2}} 
 +\ \text{terms vanishing in the limit}\ \epsilon\to0  \,,
\eal
Computing the integral, we get
\bal
F_w^b &=- \frac{ | w|  K_1(\cJ | w| )}{\pi w^2}\,.
\eal

Thus, the massive contribution is given by
\bal\label{Emassivem2}
E_m&=  2\sum_{w\neq0} \left(F_w^b   - F_w^f\right)= - 2 \sum_{w\neq0} \left(1   - e^{i w\mu}\right)\frac{ | w|  K_1(\cJ | w| )}{\pi w^2}
\\
&= - 8 \sum_{w=1}^\infty \sin^2\left({ w\mu\ov2}\right)\frac{  K_1(\cJ  w )}{\pi w}
\,,
\eal
In the large $\cJ$ limit the main contribution comes from the $w=1$ term, and we get
\bal\label{Emassivem3}
E_m&=-4\sin^2(\tfrac{\mu }{2})\,\sqrt{\frac{2}{\pi }}\,\frac{ e^{-\cJ} }{\sqrt{\cJ}} + \cO\left(e^{-J}/\cJ^{3/2}\right)\,,
\eal
which is is of the expected form.

Combining the massless and massive contributions, we obtain the GSE for large $\cJ$
\bal\label{Emasslessplusmass}
E&\approx-\frac{\mu ^2}{\pi\cJ }+\frac{|\mu+\mu'| + |\mu-\mu'| -2|\mu'| }{\cJ }-4\sin^2(\tfrac{ \mu}{2})\,\sqrt{\frac{2}{\pi }}\,\frac{ e^{-\cJ} }{\sqrt{\cJ}}\,.
\eal

\section{TBA GSE: Small Twist}\label{Sec:Small_mu}

In this section we use the \adso TBA equations, see App.\,\ref{Apx:TBASystem},  to calculate the GSE for  small twist  $\mu$ with the light-cone momentum $L$ and effective string tension $h$ kept  fixed. The result obtained  is also valid in the double-scaling limit $h\to\infty, L\to\infty$ with $\cJ\equiv L/h$ kept fixed where we can compare it with the semi-classical string computation in the previous section. Note that the TBA equations depend only on one twist $\mu$. Whether and how to introduce a  twist $\mu'$ depends on the number of massless modes~$N_0$. For $N_0=2$ we could twist 
\begin{equation}
    \log(1+Y_0^{(\dot{\alpha})})
    \qquad\to\qquad
    \begin{cases}
        \log(1+e^{ +i\mu'}Y_0^{(\dot{1})})
\,,\\
        \log(1+e^{ -i\mu'}Y_0^{(\dot{2})})\,.
    \end{cases}
\end{equation}
We will see however that the comparison of the GSE suggests $N_0=1$, which makes it hard to introduce~$\mu'$ in the TBA equations.

In order to be able to solve the $ AdS_{3} $ TBA system in the vicinity of small chemical potential $ \mu $, we need to identify the correct perturbative behaviour for the corresponding $ Y $-functions. For this purpose we recall that a neighbourhood of  BPS vacuum must be considered. At the same time vanishing twist parameter implies that ground state energy
\begin{equation}
	E_{\,\tx{BPS}} \equiv E(\mu = 0, L) = 0\,,
\end{equation}
and, therefore, from \eqref{GSE_Complete} and the TBA equations \eqref{TBA_4} and \eqref{TBA_5} for $Y_\pm$-functions the BPS vacuum $Y$-functions follow
\begin{equation}\label{BPS_vacuum}
	\mu =0 : \qquad Y_{Q} = \overline{Y}_{Q} = Y_{0}^{(\dot{\alpha})} = 0 \, , \qquad Y_{\pm}^{(\alpha)} = 1\,.
\end{equation}
However, as could be noticed, in the case of  BPS vacuum the TBA equations become divergent for $ Y_{Q}/\overline{Y}_{Q}/Y_{0}^{(\dot{\alpha})} $ functions. Hence one consistent way to regulate this, is to consider expansion around the BPS vacuum and take $ \mu \rightarrow 0 $ limit afterwards. 

\paragraph{$ \boldmath{Y} $-Ansatz and TBA solution} In this regard, it is necessary to construct some $ Y $-ansätze that would be consistent with \eqref{BPS_vacuum} and lead to a closure of the resulting system. Hence we begin by analysing the behaviour of the massive/massless modes described by $ Y_{Q/\overline{Q}/0} $-functions. From the equation for left particles, we can evaluate the structure of leading contributing terms
\begin{equation}\label{YQ}
	\begin{aligned}
		-\log Y_Q & =L \widetilde{\mathcal{E}}_Q-\log \left(1+Y_{Q^{\prime}}\right) \star K_{\fk{sl}(2)}^{Q^{\prime} Q} \\
		& -\log \left(1+\overline{Y}_{Q^{\prime}}\right) \star \widetilde{K}_{\fk{su}(2)}^{Q^{\prime} Q}  -\sum_{\dot{\alpha}=1}^{N_0} \log \left(1+Y_0^{(\dot{\alpha})}\right) \check{\star} K^{0 Q} \\
		& -\sum_{\alpha=1,2} \log \left(1-\frac{e^{i \mu_\alpha}}{Y_{+}^{(\alpha)}}\right) \hat{\star} K_{+}^{y Q}-\sum_{\alpha=1,2} \log \left(1-\frac{e^{i \mu_\alpha}}{Y_{-}^{(\alpha)}}\right) \hat{\star} K_{-}^{y Q}\,,
	\end{aligned}
\end{equation}
where $ \mu_{\alpha} = (-1)^{\alpha} \mu \, $, $ \alpha = \{ 1, 2 \} $ and $ K^{ab} $ are kernels in the appropriate mirror particle sector. The mirror energy $ \widetilde{\cl{E}}_{Q} $ for massive particles is given by
\begin{equation}\label{EQ}
	\widetilde{\cl{E}}_{Q} = \log \frac{x\left( u - i \frac{Q}{h} \right)}{x\left( u + i \frac{Q}{h} \right)} = 2 \, \tx{arcsinh}\left( \dfrac{\sqrt{\widetilde{p}^{2} + Q^{2}}}{2 h} \right)\,,
\end{equation}
where one can either use the $u$-plane rapidity or  mirror momentum $ \widetilde{p} $ of a $ Q $-particle as an independent variable.  More on mirror sectors, analytic structure and relations can be found in App. \ref{Apx:TBASystem}, \ref{Apx:S_K_Relations}.

Since for $\mu=0$ the
auxiliary $ Y $-functions are equal to 1, it is natural to assume that they admit a power series expansion  in $ \mu $
\begin{equation}\label{Y_i}
	Y_{\pm}^{(\alpha)} = 1 + B_{\pm}^{(\alpha)} \mu + C_{\pm}^{(\alpha)} \mu^2 + \cl{O}[\mu^{3}] \,.
\end{equation}
Then, taking into account that $ Y_{Q/\overline{Q}/0} $-functions vanish for $\mu=0$,  it follows immediately from \eqref{YQ} that for small $\mu$
\begin{equation}
	-\log Y_Q = -2 \log \mu \,\hat{\star}\, K_{-}^{y Q} -2 \log \mu \,\hat{\star}\, K_{+}^{y Q}+L \widetilde{\mathcal{E}}_Q + \fk{R}\,,
\end{equation}
where $ \fk{R} $ denotes the finite remainder which depends on $B_{\pm}^{(\alpha)}$ but is independent of $C_{\pm}^{(\alpha)}$. We see that the linear coefficients $ B_{\pm} $ in \eqref{Y_i} do not affect the leading $\log\mu$ dependence of $\log Y_Q$. Taking into account that $1\,\hat{\star}\, K_{\pm}^{y Q} = \dfrac{1}{2}$,  one finds that the left $ Y_Q $-function must scale quadratically in $ \mu 
$\begin{equation}\label{YQsmallmu}
	Y_Q \simeq \mb{C}_{Q} \,e^{-L \widetilde{\mathcal{E}}_Q}\, \mu^{2} +\cl{O}[\mu^{3}]\,,
\end{equation}
where $\mb{C}_{Q}$ may be a function of $\widetilde p$ (or $u$) depending on $B_{\pm}^{(\alpha)}$.

 It is then straightforward to proceed with  the right $ \overline{Y}_{Q} $ sector reaching the same conclusion.  The massless equation develops analogous structure, although in this case, attention needed for the auxiliary-massless scattering
\begin{equation}\label{YPSys}
	-\log Y_{0}^{(\dot{\alpha})} \approx L \tilde{\cl{E}}_{0} - \sum_{\alpha}\log \left(1-\frac{e^{i \mu_{\alpha} }}{Y_{-}^{(\alpha)}}\right) \hat{\star} K^{y0} - \sum_{\alpha}\log \left(1-\frac{e^{i \mu_{\alpha} }}{Y_{+}^{(\alpha)}}\right) \hat{\star} K^{y0}\,.
\end{equation}
One can easily check that the kernels $K^{y0}$ satisfy the relation $1\,\hat{\star}\, K_{\pm}^{y 0} = \dfrac{1}{2}$, and therefore the small $\mu$ behaviour of $Y_{0}^{(\dot{\alpha})}$ is given again by \eqref{YQsmallmu} with $Q=0$.

Having determined the small $\mu$ behaviour of $Y_Q$, $ \overline{Y}_{Q} $ and $ Y_{0}^{(\dot{\alpha})} $ functions, we can use the TBA equations for $Y_{\pm}^{(\alpha)} $ functions to find 
the linear coefficients $ B_{\pm}^{(\a)} $. We have for $Y_{+}^{(\alpha)}$
\begin{equation}
	\begin{aligned}
		\log Y_{+}^{(\alpha)} & = \log \left(1+\overline{Y}_Q\right) \star K_{-}^{Q y}  - \log \left(1+Y_Q\right) \star K_{+}^{Q y} 
		 \\
   &\qquad\qquad\qquad\qquad\qquad\qquad- \sum_{\dot{\alpha} = 1}^{N_0} \log \left( 1+Y_0^{(\dot{\alpha})} \right) \check{\star} K^{0 y} \,,
	\end{aligned}
\end{equation}
and a similar equation for $Y_{-}^{(\alpha)}$.
Taking into account the leading order behaviour of the massive/massless $Y$-functions, we see immediately that $ B_{\pm}^{(\a)} =0$, and therefore  the finite remainder $ \fk{R} $ is equal to 0 for all 
$Y_Q,  \overline{Y}_{Q}, Y_{0}^{(\dot{\alpha})} $ functions.

Thus, the leading order solution of the TBA equations for small $\mu$ is given by 
\begin{equation}\label{TBAsol1}
	Y_Q=\overline Y_Q \approx \mu ^2 e^{-L \tilde{\cl{E}}_{Q}}\,, \qquad Y_{0}^{(1, \, 2)} \approx \mu ^2 e^{-L \tilde{\cl{E}}_{0}}\,,\qquad Y_{\pm}^{(\alpha)}\approx 1\,.
\end{equation}
The solution is similar to the one for the $AdS_5$ case  \cite{Frolov:2009in} where one finds  $Y_Q \approx \mu ^2 4Q^2e^{-L \tilde{\cl{E}}_{Q}}$, $Y_{\pm}^{(\alpha)}\approx 1$. The extra $Q^2$ dependence of $Y_Q$ appears because the dimension of a bound state representation is $16Q^2$.

\paragraph{Ground State Energy.} 
We can now obtain the ground state energy by expanding \eqref{GSE_Complete} to the first order in $Y$-functions and using the solution \eqref{TBAsol1}
\begin{equation}\label{E_cp_small}
	\begin{aligned}
		E(h, L,\mu) & \approx - \mu^2 \left[ \dfrac{1}{2 \pi} \int\displaylimits_{-\infty}^{+ \infty} d \widetilde{p}\, N_0\, e^{-L \tilde{\cl{E}}_{0}}+ \sum_{Q = 1}^{+\infty} \, \dfrac{1}{2 \pi}\int\displaylimits_{-\infty}^{+ \infty} d \widetilde{p}\, 2 e^{-L \tilde{\cl{E}}_{Q}}  \right] \\
		& = -N_0\frac{\mu^2}{\pi} \frac{4 h L}{4 L^2-1}-\frac{\mu^2}{\pi} \cl{I}(h, L) \,.
			\end{aligned}
\end{equation}
Here the first term on the second line is the contribution of massless particles and the massless integral in \eqref{E_cp_small} can be computed analytically for $ L > \dfrac{1}{2} $.  The second term denotes the contribution of massive particles with  the function  $  \cl{I}$ given by 
\bal\label{IhL3}
\cl{I}(h,L) &= \sum_{Q = 1}^{+\infty} Q\int\displaylimits_{-\infty}^{\infty} \mathrm{d} p\,  e^{-2L\,\arcsinh\left( \dfrac{Q}{2 h}\sqrt{p^{2} + 1} \right)} 
\\
&=  \sum_{Q = 1}^{+\infty} Q\int\displaylimits_{-\infty}^{\infty} \mathrm{d} p\,  \left(\sqrt{\frac{\left(p^2+1\right) Q^2}{4
		h^2}+1}+ \sqrt{\frac{\left(p^2+1\right) Q^2}{4
		h^2}}\right)^{-2 L}\,,
\eal
where in the original integral we rescaled the mirror momentum $ \widetilde{p} $ as $ \widetilde{p}  = Q\,p$.

By analysing the massive part, one can note that it can be transformed into a double sum and further resummation over $ Q $ can be performed leading to the following power series in $h$ for  $\cl{I}(h,L)$ 
\begin{equation}\label{S_Massive}
\cl{I}(h,L)  = L\,\sum_{k=0}^\infty(-1)^k  \frac{  \Gamma
		\left(k+L-\frac{1}{2}\right) \Gamma
		\left(k+L+\frac{1}{2}\right)}{\Gamma (k+1) \Gamma (k+2 L+1)}\zeta(2 k+2 L-1)(2h)^{2 k+2 L}\,,
\end{equation}
where the convergence of the sum over $Q$, and the sum over $h$ requires 
\begin{equation}
	\cl{I}_{\tx{conv}}: \qquad L > 1 \, , \,\,\, |h| < \dfrac{1}{2}\,.
\end{equation} 
Let us also mention that 
in the \ads case the function $\cI_{AdS_5}$ is defined similarly to \eqref{IhL3} but the summand contains an extra factor of $Q^2$: 
\bal
\cI_{AdS_5}(h,L)\equiv  \sum_{Q = 1}^{+\infty} Q^2 \int_{-\infty}^{\infty} \mathrm{d}  \widetilde{p}\, e^{-L \tilde{\cl{E}}_{Q}}\,.
\eal
Its expansion in powers of $h$ is given by \eqref{S_Massive} with one replacement: $\zeta(2 k+2 L-1)\to \zeta(2 k+2 L-3)$. As a result, it is convergent for $L>2$. Note also that the GSE in the $AdS_5$ case is given by $E_{AdS_5}(h,L)=-\frac{2\mu^2}{\pi} \cl{I}_{AdS_5}(h, L)$.

\paragraph{Limit $h\to\infty, L\to\infty$ with $L/h$  fixed.}
The GSE \eqref{E_cp_small} has been derived under an assumption $\mu\ll1$ with $h,L$ fixed. The solution \eqref{TBAsol1}, however, is also valid in the scaling limit $h\to\infty, L\to\infty$ with $\cJ\equiv L/h$ kept fixed where \eqref{TBAsol1} simplifies to 
\begin{equation}\label{TBAsol1Lhinfty}
	Y_Q=\overline Y_Q \approx \mu ^2 e^{-\sqrt{\widetilde{p}^2+Q^2}\cJ}\,, \qquad Y_{0}^{(1, \, 2)} \approx \mu ^2e^{-|\widetilde{p}|\cJ}\,,\qquad Y_{\pm}^{(\alpha)}\approx 1\,.
\end{equation}
Then we get
\bal
\cl{I}(\cJ )&=  \sum_{Q = 1}^{+\infty} Q\int\displaylimits_{-\infty}^{\infty} \mathrm{d} p\,  e^{-\cJ\,Q\sqrt{p^{2} + 1} } 
= \int\displaylimits_{-\infty}^{\infty} \mathrm{d} p\,  {1\ov 4 \sinh ^2\left(\frac{1}{2} \cJ \sqrt{p^2+1}\right)}\,.
\eal
For finite $\cJ$ the integral cannot be computed analytically but for large $\cJ$ one gets
\bal\label{IcJJ1}
\cl{I}(\cJ\gg1 )&\approx \frac{\sqrt{\frac{\pi }{2}}}{(\cosh \cJ-1) \sqrt{\cJ \coth
		\frac{\cJ}{2}}} \to 
\frac{\sqrt{2\pi} e^{-\cJ}}{ \sqrt{\cJ }} \,.
\eal
Thus, the GSE becomes
\begin{equation}\label{E_cp_smallLhinfty}
	\begin{aligned}
		E(\cJ\gg 1) &  \approx -\frac{\mu^2}{\pi}\left( \frac{N_0}{\cJ}+\frac{\sqrt{2\pi} e^{-\cJ}}{ \sqrt{\cJ }} \right)\,.
			\end{aligned}
\end{equation}
Obviously,  massless particles provide the leading contribution  of order $1/\cJ$ while, as expected, the contribution of massive particles is exponentially suppressed for large $\cJ$.  Comparing \eqref{E_cp_smallLhinfty} with \eqref{Emasslessplusmass}, we see that to get an agreement one has to set $N_0$ to 1 and choose $\mu'=\mu$.

Massive particles however contribute more to the GSE for small $\cJ$ where
 one finds
\bal\label{IcJJ2}
\cl{I}(\cJ\ll1 )&\approx {\pi\ov \cJ^2}-{1\ov \cJ}\,,
\eal
and
\begin{equation}\label{E_cp_smallLhinfty2}
	\begin{aligned}
		E(\cJ\ll 1) &\approx -\frac{\mu^2}{\pi}\left( \frac{N_0-1}{\cJ}+ {\pi\ov \cJ^2} \right)\,.
			\end{aligned}
\end{equation}
One see that for small $\cJ$ massive particles contribute to  order $1/\cJ$, and  for $N_0=1$ completely compensate the  massless contribution.

 It is interesting to compare these formulae with the ones for the $ AdS_{5} $ superstring where the $ \cl{J} $-dependent integral arises in the form
\begin{equation}\label{IcJ5}
\begin{aligned}
	\cl{I}_{AdS_{5}}(\cl{J}) &= \int\displaylimits_{-\infty}^{+\infty} \dd p \, \dfrac{2 \sinh^{2}\left(\frac{J}{2} \sqrt{p^2+1} \right)+3}{8 \sinh^{4}\left(\frac{J}{2} \sqrt{p^2+1} \right)}\\
 &=\cl{I}(\cl{J}) +\int\displaylimits_{-\infty}^{+\infty} \dd p \, \dfrac{3}{8 \sinh^{4}\left(\frac{\cJ}{2} \sqrt{p^2+1} \right)} \,.
\end{aligned}
\end{equation}
The second term is clearly subleading for $\cJ\gg1$, and one finds
\begin{equation}
	\begin{aligned}
		E_{AdS_{5}}(\cJ\gg 1) &  = -\frac{2\mu^2}{\pi} \frac{\sqrt{2\pi} e^{-\cJ}}{ \sqrt{\cJ }} \,.
			\end{aligned}
\end{equation}
On the other hand for small $\cJ$ the second term in \eqref{IcJ5} scales as $ \cl{J}^{-4} $ and provides the main contribution 
\begin{equation}\label{E_cp_smallLhinfty3}
	\begin{aligned}
\cl{I}_{AdS_{5}}(\cl{J} \ll 1) = \dfrac{3 \pi}{\cl{J}^{4}}\,,\qquad E_{AdS_{5}}(\cJ\ll 1) &  = -\frac{6\mu^2}{\cl{J}^{4}} \,.
			\end{aligned}
\end{equation}
It is unclear to us what is a reason for such a different behaviour of the GSE in the $AdS_3$ and $AdS_5$ cases.

\paragraph{Regime $ \mu \ll 1 $, $\mu\, h^L\ll 1$,  $h\gg L$ with $L$  fixed.}

Finally,  we can consider the regime where $h$ is much larger than $L$ which is kept fixed. Then, we have to assume that $\mu\, h^L\ll 1$ otherwise $Y$-functions would be large.\footnote{For finite $\mu$ and $L$ we expect the ground state to become unstable, and tachyonic modes to appear, see \cite{Bajnok:2013wsa} for a related discussion.} In this case 
 the sum over $Q$ in \eqref{IhL3} can be replaced with an integral, and we get
\bal
\cl{I}(h\gg L,L)&= \sum_{Q = 1}^{+\infty} Q\int\displaylimits_{-\infty}^{\infty} \mathrm{d} p\,  \left(\sqrt{\frac{\left(p^2+1\right) Q^2}{4
		h^2}+1}+ \sqrt{\frac{\left(p^2+1\right) Q^2}{4
		h^2}}\right)^{-2 L}
\\
&\approx h^2\int_0^\infty \mathrm{d}q\,q\int\displaylimits_{-\infty}^{\infty} \mathrm{d} p\,  \left(\sqrt{\frac{\left(p^2+1\right) q^2}{4}+1}+ \sqrt{\frac{\left(p^2+1\right) q^2}{4}}\right)^{-2 L} \,.
\eal
The integrals can be easily taken by making the substitution
\bal
q={2\sinh x\ov \sqrt{p^2+1}}\,,
\eal
and we obtain
\bal
\cl{I}(h\gg L,L)&\approx h^2{\pi\ov L^2-1} \,,
\eal
and
\begin{equation}\label{Ehlarge}
	\begin{aligned}
		E(h\gg L, L) \approx -\frac{\mu^2}{\pi}\left(N_0 \frac{4 h L}{4 L^2-1}+{h^2\pi\ov L^2-1}\right) \,.
			\end{aligned}
\end{equation}
In this case massless modes provide a subleading contribution.

Computing a similar integral for the \ads case, one gets 
\bal
\cl{I}_{AdS_5}(h\gg L,L)&\approx h^4\frac{3 \pi }{L^4-5 L^2+4}\,,\\
E_{AdS_5}(h\gg L, L) &\approx -\frac{\mu^2}{\pi}\frac{6 h^4 \pi }{L^4-5 L^2+4} \,.
\eal
We see again that there is a substantial difference between the \adso and \ads superstrings.

\section{TBA GSE: Finite Twist}\label{Sec:LCA}

\paragraph{Large $L \tilde{\cl{E}}_{Q}$.}
One can easily see from the form of the TBA equations \ref{Apx:TBASystem}  that 
$Y_Q,  \overline{Y}_{Q}, Y_{0}^{(\dot{\alpha})} $ functions are small if $L \tilde{\cl{E}}_{Q}$ are large.  This clearly happens either if $L$ is large with $h$ fixed or $h$ is small with $L$ fixed while the value of the twist $\mu$ is irrelevant. Small parameters then become  $e^{-L \tilde{\cl{E}}_{Q}}$, and repeating the analysis in the previous section, one gets the leading order solution 
\begin{equation}\label{TBAsolL}
	Y_Q=\overline Y_Q \approx 4 \sin ^2(\tfrac{\mu }{2})\, e^{-L \tilde{\cl{E}}_{Q}}\,, \qquad Y_{0}^{(1, \, 2)} \approx 4 \sin ^2(\tfrac{\mu }{2})\, e^{-L \tilde{\cl{E}}_{0}}\,,\qquad Y_{\pm}^{(\alpha)}\approx 1\,,
\end{equation}
which differs from \eqref{TBAsol1} just by the replacement $\mu^{2}  \rightarrow  4 \sin ^2\frac{\mu }{2}$, and therefore the GSE is given by
\begin{equation}\label{E_Large_L}
	E(h,L,\mu) \approx - \frac{4}{\pi} \sin^2(\tfrac{\mu }{2}) \left( N_0 \frac{4 h L}{4 L^2-1} +\cl{I}(h,L)\right)\,.
\end{equation}
For small $h$  with $L$ fixed we can use \eqref{S_Massive}, and get
\begin{equation}\label{Esmallh}
	E(h\ll1,L,\mu) \approx - \frac{4}{\pi} \sin^2(\tfrac{\mu }{2}) \left( N_0 \frac{4 h L}{4 L^2-1} +\frac{\sqrt{\pi }  \Gamma
   \left(L-\frac{1}{2}\right)}{\Gamma (L)}\zeta (2 L-1)h^{2 L}\right) \,.
\end{equation}
The contribution of massive particles appears only at order $h^{2L}$ while massless particles begin to contribute already at the linear order. It is interesting to note that due to massless particles contributions the expansion is in powers of $h$ while in the \ads case it is in powers of $h^2$.

 For large $L$ with $h$ fixed the main contribution of massive particles comes from $Q=1$,  and the constant and quadratic in $\widetilde p$ terms in the small $\widetilde p$ expansion of $\widetilde{\cl{E}}_{1}$. Performing the expansion and computing the integral, we get
\bal\label{IhLL}
\cI(h,L\gg h)&\approx \frac{\sqrt{\pi } \sqrt[4]{4 h^2+1} \left(\sqrt{\frac{1}{4
			h^2}+1}+\frac{1}{2 h}\right)^{-2 L}}{\sqrt{L}} \,,
\eal
\begin{equation}\label{ElargeL}
	E(h,L\gg h,\mu) \approx - \frac{4}{\pi} \sin^2(\tfrac{\mu }{2}) \left( N_0 \frac{h}{L} + \frac{\sqrt{\pi } \sqrt[4]{4 h^2+1} \,(2h)^{2L}}{\sqrt{L}\,\left(\sqrt{4
			h^2+1}+1\right)^{2 L}}\right) \,.
\end{equation}
Just as for the case of small $h$  with $L$ fixed the leading contribution comes from  the massless particles  and is proportional to $h/L$. It is also clear that for any $h$ the massive contribution is exponentially suppressed. 

For small and large $h$ the formula \eqref{IhLL}  simplifies
\bal\label{ElargeLh}
\cI(h\ll1,L\gg h)&\approx \sqrt{\frac{\pi} {L}}\, h^{2 L} \,,\\
\cI(h\gg1,L\gg h)&\approx \sqrt{2\pi \frac{h}{L}}\, e^{- {L\ov h}} =\sqrt{ \frac{2\pi}{\cJ}}\, e^{- \cJ} \,,
\\
E(h\gg1,L\gg h,\mu) &\approx - \frac{4}{\pi} \sin^2(\tfrac{\mu }{2}) \left( N_0 \frac{1}{\cJ} + \sqrt{ \frac{2\pi}{\cJ}}\, e^{- \cJ}\right) \,,
\eal
and the energy agrees with \eqref{Esmallh} and \eqref{IcJJ1}.

Let us stress that for $\mu=\pi$ formulae \eqref{Esmallh} and \eqref{ElargeL}  should provide us with the GSE   of the RR \adso light-cone string theory in the odd-winding number sector with anti-periodic fermions and nonsupersymmetric vacuum.

\paragraph{Twisted Generalised L\"uscher formula.}

We have found that both small $ \mu $ and asymptotically large $ L \tilde{\cl{E}}_{Q}$ regimes share extra contribution that comes from the massless sector and take a very simple and compact form.
The massless contribution does not vanish exponentially for large $L$, and it is interesting to check whether it is
consistent with  generalised L\"uscher formula \cite{Luscher:1985dn,Janik:2007wt,Bajnok:2008bm,Hatsuda:2008na,Ahn:2011xq,Bombardelli:2013yka} which in the presence of the twist 
takes the form
\begin{equation}\label{E_TD_Luescher}
	\begin{aligned}
		E(h, L) & \approx -\sum_{\dot\a = 1}^{N_0} \int\displaylimits_{-\infty}^{\infty}  \frac{d \widetilde{p}}{2 \pi}  e^{-L \tilde{\cl{E}}_{0}} \fk{F}_{0}^{(\dot\a)}  - \sum_{Q = 1}^{+\infty} \, \int\displaylimits_{-\infty}^{\infty} \frac{d \widetilde{p}}{2 \pi}  e^{-L \tilde{\cl{E}}_{Q}} (\fk{F}_{Q}+\bar{\fk{F}}_{Q} )\,,
	\end{aligned}
\end{equation}
where
\begin{equation}
	\fk{F}_{X} = \tx{Tr}_{X} e^{i (\pi + \mu)F} \,,
\end{equation}
and $ \tx{Tr}_{X} $ is over appropriate  representations $ X = \{ Q/\overline{Q}/0 \} $. For the \adso superstring all the representations are four-dimensional, and moreover each of them is the tensor product of two two-dimensional  representations: $X=X'\otimes X''$. Each two-dimensional representation has one boson and one fermion, and $ F = F'-F'' $ plays the role of a fermion number operator of the corresponding four-dimensional representation where $F'$, $F''$ are fermion number operators of the left/right two-dimensional representations. 

Calculating the trace, one gets for all the representations
\begin{equation}
\begin{aligned}
 \tx{Tr}_{X} e^{i (\pi + \mu) F} &=\left(\tx{Tr}_{X'} e^{i (\pi + \mu) F'}\right) \left(\tx{Tr}_{X''} e^{-i (\pi + \mu) F''}\right)\\
 &=  \left(1-e^{i \mu }\right) \left(1-e^{-i \mu }\right) = 4 \sin^{2} (\tfrac{\mu }{2}) \,.
 \end{aligned}
\end{equation}
Clearly,  the twisted generalised Lüscher expression \eqref{E_TD_Luescher} is in full agreement with the TBA formula \eqref{E_Large_L}. 

\section{Twisted mixed-flux \texorpdfstring{\boldmath\adso}{AdS3xS3xT4} superstring}\la{mixedflux}
In this section we first generalise the semiclassical consideration in section \ref{lcstring} to the twisted lightcone string on the  mixed-flux  \adso background. Then, we conjecture that up to unknown dressing factors the TBA equations for the mixed-flux  \adso superstring are similar to the pure RR ones, and therefore in the situation where the $Y_Q$-functions are small a solution takes the same form. The only difference is that one uses the mirror dispersion relation of the mixed-flux theory. 
In fact, as far as the semiclassical computation is concerned, we can consider the even more general one-parameter family of $AdS_3\times S^3\times S^3\times S^1$  backgrounds (where the parameter is the relative radius of the two spheres). It  reduces to the  mixed-flux  \adso superstring when one sphere blows up, and  the corresponding untwisted lightcone string sigma model was considered in \cite{Dei:2018yth}. For this one-parameter family of backgrounds, however, the mirror TBA is not known (and very little is known aside from the matrix structure of  the S-matrix~\cite{Borsato:2015mma} and BPS spectrum~\cite{Baggio:2017kza}).  

\subsection*{Lightcone \texorpdfstring{\boldmath $AdS_3\times S^3\times S^3\times S^1$}{AdS3xS3xS3xS1} string GSE}

The quadratic action of the lightcone $AdS_3\times S^3\times S^3\times S^1$ string derived in \cite{Dei:2018yth} takes the form
\bal
S = \int \dd \tau\int_0^\cJ \dd \sigma \, \cl{L}\,,\qquad	\cl{L} = \cl{L}^b + \cl{L}^f\,,
\eal
where $\cl{L}^b$ and $\cl{L}^f$ are the bosonic and fermionic quadratic Lagrangians, respectively
\begin{equation}\label{lagrbf}
	\begin{aligned}
		\cl{L}^b & =\sum_{j=1}^4{1\ov2}\left[ \left|\partial_k z_j\right|^2 - m_{b,j}^2 \left| z_j\right|^2 +i q m_{b,j}(\bar z_j z_j' - \bar z_j' z_j) \right] \,, \\[1ex]
		\cl{L}^f &  =\sum_{j=1}^4\Big[ i\bar\theta_{1j} \left(\dot\theta_{1j}-i\hat q \theta_{2j}' +q\theta_{1j}'\right)  \\
  &\qquad\qquad\qquad +i\bar\theta_{2j} \left(\dot\theta_{2j}+i\hat q \theta_{1j}' -q\theta_{2j}'\right) -m_{f,j} (\bar\theta_{1j} \theta_{1j}-\bar\theta_{2j} \theta_{2j})\Big]\,.
	\end{aligned}
\end{equation}
Here  $\hat q = \sqrt{1-q^2}$, and the masses are given by
\bal\la{massb}
m_{b,1}=1\,,\quad m_{b,2}=\cos \varphi\cos\w\,,\quad m_{b,3}=\sin \varphi\sin\w\,,\quad m_{b,4}=0\,,
\eal
for the bosons, while for the fermions
\bal\la{massf}
m_{f,1}&={1+\cos(\varphi-\w)\ov2}\,,\quad m_{f,2}={1+\cos(\varphi+\w)\ov2}\,,\quad 
\\
m_{f,3}&={1-\cos(\varphi+\w)\ov2}\,,\quad m_{f,4}={1-\cos(\varphi-\w)\ov2}\,,
\eal
where $\varphi$ parametrises the $AdS_3\times S^3\times S^3\times S^1$ background and is related to the commonly used parameter $\a$ as $\a = \cos^2\varphi$, and $\w$ parametrises the one-parameter family of choice of gauge-fixing that determines the lightcone string sigma models. The models are supersymmetric only for $\w=\pm\varphi$. Then, the \adso model is recovered at $\w=\varphi=0$. For any values of $\varphi,\w$ the masses satisfy the following important relation
\bal\la{mass2}
\sum_{j=1}^4 m_{b,j}^2 = \sum_{j=1}^4 m_{f,j}^2 \,,
\eal
which as we will see in a moment allows one to introduce a regularisation leading to a finite ground state energy of the lightcone model.

We impose the most general twisted boundary conditions
\bal
z_j(\sigma + \cJ,\tau) &= e^{i \mu_j^b}z_j(\sigma,\tau)\,,\quad \theta_{a,j}(\sigma + \cJ,\tau) &= e^{i \mu_j^f}\theta_{a,j}(\sigma,\tau)\,,\quad a=1,2\,,
\eal
and find the bosonic and fermionic frequencies
\bal
\w_{n,j}^b &= \sqrt{\left({2\pi\ov\cJ}(n+\hat\mu_j^b)+qm_{b,j} \right)^2 +\hat q^2m_{b,j}^2 }\,,\\
\w_{n,j}^f &= \sqrt{\left({2\pi\ov\cJ}(n+\hat\mu_j^f)+qm_{f,j} \right)^2+\hat q^2m_{f,j}^2 }\,.
\eal
Then, the GSE is given by
\bal
E = \sum_{j=1}^4\sum_{n=-\infty}^\infty \w_{n,j}^b - \sum_{j=1}^4\sum_{n=-\infty}^\infty \w_{n,j}^f\,.
\eal
We use the following regularisation
\bal
E = \sum_{j=1}^4\sum_{n=-\infty}^\infty \w_{n,j}^b\,\exp\left({-{\e\ov \hat qm_{b,j}}\w_{n,j}^b}\right) - \sum_{j=1}^4\sum_{n=-\infty}^\infty \w_{n,j}^f\,\exp\left({-{\e\ov \hat qm_{f,j}}\w_{n,j}^f}\right) \,.
\eal
which is a generalisation of \eqref{Emassive2}. We also assume that all the masses are nonzero.

By using again the Poisson summation formula, we get
\bal\label{Eads3s3s3s1}
E&= \sum_{j=1}^4\sum_{w=-\infty}^\infty \left(F_{w,j}^b   - F_{w,j}^f\right)= \sum_{j=1}^4\left(F_{0,j}^b   - F_{0,j}^f\right)+  \sum_{j=1}^4\sum_{w\neq0} \left(F_{w,j}^b   - F_{w,j}^f\right)
\,.
\eal
Here
\bal
F_{w,j}^\star &=  \int_{-\infty}^\infty dn\, \w_{n,j}^\star\,\exp\left({-{\e\ov \hat qm_{\star,j}}\w_{n,j}^\star}\right) e^{-i2\pi wn}  
\\
&=e^{i w(\mu_j^\star+qm_{\star,j}\cJ)}\frac{\cl{J}\hat q^2m_{\star,j}^2}{2 \pi}\int_{-\infty}^\infty dx\, \sqrt{x^2 + 1}e^{-\epsilon\, \sqrt{x^2 + 1}} e^{-i \hat qm_{\star,j}\cJ wx}\,,
\eal
where $\star = b,f$.

Taking into account the relation \eqref{mass2}, we see that $\sum_{j=1}^4\left(F_{0,j}^b   - F_{0,j}^f\right)=0$, and therefore we only need to calculate $F_w^\star$ for $w\neq 0$. This has been done in section \ref{lcstring}, and by using the results we find the GSE 
\bal\label{Eads3s3s3s1a}
E=-2\hat q\, \sum_{j=1}^4\sum_{w=1}^\infty  &\left(m_{b,j}\cos\left( w(\mu_j^b+qm_{b,j}\cJ)\right) \frac{    K_1(\hat qm_{b,j}\cJ  w )}{\pi w} \right.
\\
&\left.  - m_{f,j}\cos\left( w(\mu_j^f+qm_{f,j}\cJ)\right) \frac{    K_1(\hat qm_{f,j}\cJ  w)}{\pi w} \right)
\,.
\eal
If some of the masses are equal to 0 then we can either use the zeta-function regularisation, or take the limit $m\to 0$ in the formula. Assuming that $-\pi<\mu<\pi$, we get for the massless contribution
\bal\label{Eads3s3s3s1m0}
E_0&=-2\hat q\, \lim_{m\to0}\sum_{w=1}^\infty (m\cos\left( w(\mu+qm\cJ)\right) \frac{    K_1(\hat qm\cJ  w )}{\pi w} 
\\
&=-\sum_{w=1}^\infty \frac{2 \cos (\mu  w)}{\pi  \cJ w^2} = -\frac{\pi }{3 \cJ}+\frac{|\mu|}{\cJ}-\frac{\mu ^2}{2 \pi  \cJ}
\,,
\eal
which agrees with the result obtained by using the zeta-function regularisation.

Thus, assuming that we have $N_0^b$ and $N_0^f$ complex massless bosons and fermions, and $N_{\rm m}^b$ and $N_{\rm m}^f$ complex massive bosons and fermions, we can write \eqref{Eads3s3s3s1a} in the form
\bal\label{Eads3s3s3s1b}
E&=-\frac{\pi(N_0^b-N_0^f) }{3 \cJ}+ \sum_{j=1}^{N_{0}^b}\left(\frac{|\mu_{0j}^b|}{\cJ}-\frac{(\mu_{0j}^b)^2}{2 \pi  \cJ}\right)- \sum_{j=1}^{N_{0}^f}\left(\frac{|\mu_{0j}^f|}{\cJ}-\frac{(\mu_{0j}^f)^2}{2 \pi  \cJ}\right)
\\
&-2\hat q\, \sum_{j=1}^{N_{\rm m}^b}\sum_{w=1}^\infty  m_{b,j}\cos\left( w(\mu_j^b+qm_{b,j}\cJ)\right) \frac{    K_1(\hat qm_{b,j}\cJ  w )}{\pi w} 
\\
&  +2\hat q\, \sum_{j=1}^{N_{\rm m}^f}\sum_{w=1}^\infty  m_{f,j}\cos\left( w(\mu_j^f+qm_{f,j}\cJ)\right) \frac{    K_1(\hat qm_{f,j}\cJ  w)}{\pi w} )
\,.
\eal
In the supersymmetric case $N_0^b=N_0^f=N_0$, $N_{\rm m}^b=N_{\rm m}^f=N_{\rm m}$, $m_{b,j}=m_{f,j}=m_{j}$, and \eqref{Eads3s3s3s1b} takes a simpler form
\bal\label{Eads3s3s3s1c}
E&= \sum_{j=1}^{N_{0}}\left(\frac{|\mu_{0j}^b|-|\mu_{0j}^f|}{\cJ}-\frac{(\mu_{0j}^b)^2-(\mu_{0j}^f)^2}{2 \pi  \cJ}\right)\\
&-4\hat q\, \sum_{j=1}^{N_{\rm m}}\sum_{w=1}^\infty  m_{j}\sin\left( w{\mu_j^b+\mu_j^f+2qm_{j}\cJ\ov2}\right)\sin\left( w{\mu_j^f-\mu_j^b\ov2}\right) \frac{    K_1(\hat qm_{j}\cJ  w )}{\pi w} 
\,.
\eal
In the large $\cJ$ limit the main contribution comes from $w=1$, and one gets 
\bal\label{Eads3s3s3s1d}
&E(\cJ\gg 1)= \sum_{j=1}^{N_{0}}\left(\frac{|\mu_{0j}^b|-|\mu_{0j}^f|}{\cJ}-\frac{(\mu_{0j}^b)^2-(\mu_{0j}^f)^2}{2 \pi  \cJ}\right)\\
&\qquad-4\hat q\, \sum_{j=1}^{N_{\rm m}} m_{j}\sin\left( {\mu_j^b+\mu_j^f+2qm_{j}\cJ\ov2}\right)\sin\left( {\mu_j^f-\mu_j^b\ov2}\right)\frac{e^{-\hat qm_{j}\cJ}}{\sqrt{2 \pi } \sqrt{\hat qm_{j}\cJ}}
\,.
\eal
We can now apply \eqref{Eads3s3s3s1d} to the mixed-flux \adso string where 
$N_0=2$, $N_{\rm m}=2$, and the twists are given by
\bal
\mu_{01}^b &= \mu+\mu'\,,\quad \mu_{02}^b = \mu-\mu'\,,\quad \mu_{01}^f = \mu'\,,\quad \mu_{02}^f = -\mu'\,,\quad
\\
\mu_{1}^b &= 0\,,\quad \mu_{2}^b = 0\,,\quad \mu_{1}^f = \mu\,,\quad \mu_{2}^f = -\mu\,,\quad m_1=m_2=1\,,
\eal
and therefore
\bal\label{Eads3s3t4d}
E_{\rm T^4}(\cJ\gg 1)=&-\frac{\mu ^2}{\pi\cJ }+\frac{|\mu+\mu'| + |\mu-\mu'| -2\mu' }{\cJ }\\
&-4\sin^2\left({ \mu\ov2}\right)\hat q\cos\left(q\cJ\right)\sqrt{\frac{2}{\pi }}\,\frac{ e^{-\hat q\cJ} }{\sqrt{\hat q\cJ}} 
\,.
\eal
Thus, the massless contribution is independent of $q$ while the massive contribution contains a highly oscillating factor. It is interesting that if $q\cJ = {\pi\ov 2}+ \pi\ell$, $\ell\in\bZ$ then the terms with $w=2$ would provide the leading contribution at large $\cJ$.

In the next subsection we will reproduce \eqref{Eads3s3t4d} from TBA equations for the mixed-flux \adso string.

\subsection*{GSE from TBA}

The mirror theory of the mixed-flux \adso string is nonunitary, and it is unclear whether one can justify using it to derive the  mixed-flux TBA equations.%
\footnote{The derivation of the TBA for such non-unitary mirror models was considered, at least formally, in~\cite{Dei:2018mfl,Dei:2018jyj} for the $q=1$ case, and successfully matched with the WZNW model~results.}
Still, one can try to engineer a set of integral equations which would determine the ground state energy of the string theory.
Since the Bethe equations and the bound states of the mixed-flux \adso string is similar to the ones of the RR \adso string, it is natural to assume that the TBA equations in the mixed-flux case take the same form as in the RR TBA system, see App.\,\ref{Apx:TBASystem}. Clearly, the dressing kernels and the energy dispersion relations would have to be replaced with the ones for the mixed-flux string, and since the dressing phases are unknown writing down mixed-flux TBA equations remains a challenging problem. Nevertheless, if  $Y_Q$-functions are small then one can expect that they are given by 
\bal
Y_Q \approx \mu^2 e^{-L\tE_Q}\,,\quad \overline Y_{Q} \approx \mu^2 e^{-L\tE_{Q}^*}\,,
\eal
where $\tE_{Q}^*$ should be complex conjugate to $\tE_{Q}$ to ensure the reality of the GSE.

For finite $h$ the mixed-flux mirror dispersion relations cannot be found in an explicit analytic form. 
However, in the semi-classical limit $h\to\infty$, $L\to\infty$, $\cJ=L/h$ fixed they take the following simple form
\bal
\tE_Q = {1\ov h}\left(\sqrt{\tp^2 + \hat q^2 Q^2} +i q Q \right)\,.
\eal
Then, the GSE is given by
\begin{equation}\label{ETBAq}
	\begin{aligned}
		E(\cJ,\mu,q) & \approx -\dfrac{1}{2 \pi} \int\displaylimits_{-\infty}^{+ \infty} d \widetilde{p}\, N_0\, Y_0- \sum_{Q = 1}^{+\infty} \, \dfrac{1}{2 \pi}\int\displaylimits_{-\infty}^{+ \infty} d \widetilde{p}\,(Y_Q+\overline Y_{Q}) \\
		& = -N_0\frac{\mu^2}{\pi\cJ} -\mu^2 \cl{I}_q(\cJ) \,,
	\end{aligned}
\end{equation}
where 
\bal
 \cl{I}_q(\cJ) &=  \sum_{Q = 1}^{+\infty} \,{1\ov\pi}\int\displaylimits_{-\infty}^{+ \infty} d \widetilde{p}\,\cos(q \cJ Q)\,e^{-\cJ\sqrt{\tp^2 + \hat q^2 Q^2} }\\
 &={1\ov\pi}\sum_{Q = 1}^{+\infty} \,\hat q Q\int\displaylimits_{-\infty}^{+ \infty} d {p}\,\cos(q \cJ Q)\,e^{-\hat q Q\cJ\sqrt{p^2 + 1} }\,,
\eal
At large $\cJ$ the main contribution comes from the $Q=1$ term, and is given by
\bal
 \cl{I}_q(\cJ\gg1) = \hat q\cos\left(q\cJ\right)\sqrt{\frac{2}{\pi }}\,\frac{ e^{-\hat q\cJ} }{\sqrt{\hat q\cJ}} \,.
\eal
This agrees with the semi-classical calculation in the previous section.

\section{Conclusion and remarks} \label{conclusion}

In the present work we have calculated the leading wrapping contribution of massless and massive particles to the ground state energy of the lightcone pure RR \adso superstring with fields subject to twisted boundary conditions by using first the semi-classical string consideration and then the \adso TBA equations with any number $N_0$ of massless $Y_0^{(\dot\a)}$ functions. The comparison of the two calculations has shown that the agreement requires $N_0=1$ contrary to the conjecture made in  \cite{Frolov:2021bwp} where $N_0$ was chosen to be equal to 2. 

The $\mu$ dependence of the massless contribution in \eqref{ElargeLh}, however, disagrees with the semi-classical string result \eqref{Emasslessplusmass} even if $\mu'=\mu$. A reason for the disagreement might be in the order-of-limits problem. In the semi-classical calculation we first take $h,L\to\infty$ with $L/h$ fixed while in the TBA consideration we take $L\to \infty$ keeping $h$ fixed. It would be important to analyse the TBA equations in the limit $h,L\to\infty$ with $L/h$ fixed. If the disagreement would be resolved then it would strongly support the TBA equations with $N_0=1$.  If the disagreement  persists it would be an indication that the $su(2)_\circ$ invariant S-matrix for massless particles introduced in \cite{Borsato:2014hja}, and  assumed to be trivial in \cite{Sundin:2014ema} on the basis of perturbative calculations, is in fact nontrivial nonperturbatively. This might be also necessary in order to introduce the second twist $\mu'$ in the TBA equations. 

 The  massless contribution  has been  shown to be proportional to $hL/ (4L^2-1)$, and the contribution of massive particles appears to be subleading in all cases where $h\ll L$. We have also checked that the generalised L\"uscher formula \cite{Luscher:1985dn,Janik:2007wt,Bajnok:2008bm,Hatsuda:2008na} leads to the same leading order results as the TBA.

It would be interesting to analyse the TBA equations at the next-to-leading order where additional nontrivial contributions start to arise due to dressing phase dependent kernels. The GSE at this order can be also found by means of the next-to-leading order L\"uscher formula \cite{Ahn:2011xq,Bombardelli:2013yka}. Comparison of the results obtained by using the TBA and the  L\"uscher formula may clarify a reason for the mismatch between semiclassical calculations and the  L\"uscher formula  found in  \cite{Abbott:2015pps}.

It would also be important to generalise the recent proposal for the Quantum Spectral Curve (QSC) of pure RR \adso superstring \cite{Ekhammar:2021pys,Cavaglia:2021eqr} to the twisted superstring, and to use the QSC to calculate the GSE. This would be a definite test on whether the QSC describes massless modes and how many of them. The results of \cite{Cavaglia:2022xld}  do not answer these questions.

A very important if not ultimate test of the TBA system is to use the contour deformation trick to derive the excited states equations and to solve them numerically for large $h$ keeping $L$ fixed. One should get an expected $\sqrt{h}$ behaviour for the excited states energy.

\section*{Acknowledgements}
We would like to thank  Jean-S\'ebastien Caux  and Tristan McLoughlin for useful related discussions.

AP is supported by German Research Foundation via the Emmy Noether program “Exact Results in Extended Holography”. 
AS acknowledges support from the European Union -- NextGenerationEU, and from the program STARS@UNIPD, under project ``Exact-Holography -- A new exact approach to holography: harnessing the power of string theory, conformal field theory, and integrable models.''

\appendix

\section{\texorpdfstring{\boldmath\adso}{AdS3xS3xT4} TBA system}\label{Apx:TBASystem}

In this appendix we list the \adso TBA equations proposed in  \cite{Frolov:2021bwp} but for generality and to simplify the comparison with the semi-classical string results we consider $N_0$  massless $Y_0^{\dot\a}$-functions.
The  {\adso} TBA system \cite{Frolov:2021bwp} contains $ Y_{\cl{X}} $-functions and convolution kernels $ K^{pq} $, and is given as follows.

The left, right and massless $Y$-functions for momentum-carrying Bethe roots satisfy the following  equations
\begin{equation}\label{TBA_1}
	\begin{aligned}
		-\log Y_Q & =L \widetilde{\mathcal{E}}_Q-\log \left(1+Y_{Q^{\prime}}\right) \star K_{\fk{sl}(2)}^{Q^{\prime} Q} \\
		& -\log \left(1+\overline{Y}_{Q^{\prime}}\right) \star \widetilde{K}_{\fk{su}(2)}^{Q^{\prime} Q}  -\sum_{\dot{\alpha}=1}^{N_0} \log \left(1+Y_0^{(\dot{\alpha})}\right) \check{\star} K^{0 Q} \\
		& -\sum_{\alpha=1,2} \log \left(1-\frac{e^{i \mu_\alpha}}{Y_{+}^{(\alpha)}}\right) \hat{\star} K_{+}^{y Q}-\sum_{\alpha=1,2} \log \left(1-\frac{e^{i \mu_\alpha}}{Y_{-}^{(\alpha)}}\right) \hat{\star} K_{-}^{y Q}\,,
	\end{aligned}
\end{equation}
\vspace{2mm}
\begin{equation}\label{TBA_2}
	\begin{aligned}
		-\log \overline{Y}_Q & =L \widetilde{\mathcal{E}}_Q-\log \left(1+\overline{Y}_{Q^{\prime}}\right) \star K_{\fk{su}(2)}^{Q^{\prime} Q} \\
		& -\log \left(1+Y_{Q^{\prime}}\right) \star \widetilde{K}_{\fk{sl}(2)}^{Q^{\prime} Q} -\sum_{\dot{\alpha}=1}^{N_0} \log \left(1+Y_0^{(\dot{\alpha})}\right) \check{\star} \widetilde{K}^{0 Q}\\
		& -\sum_{\alpha=1,2} \log \left(1-\frac{e^{i \mu_\alpha}}{Y_{+}^{(\alpha)}}\right) \hat{\star} K_{-}^{y Q}-\sum_{\alpha=1,2} \log \left(1-\frac{e^{i \mu_\alpha}}{Y_{-}^{(\alpha)}}\right) \hat{\star} K_{+}^{y Q} \,,
	\end{aligned}
\end{equation}
\vspace{2mm}
\begin{equation}\label{TBA_3}
	\begin{aligned}
		-\log Y_0^{(\dot{\alpha})} & =L \widetilde{\mathcal{E}}_0-\sum_{\dot{\beta}=1}^{N_0} \log \left(1+Y_0^{(\dot{\beta})}\right) \check{\star} K^{00} \\
		& -\log \left(1+Y_Q\right) \star K^{Q 0}-\log \left(1+\overline{Y}_Q\right) \star \widetilde{K}^{Q 0} \\
		& -\sum_{\alpha=1,2} \log \left(1-\frac{e^{i \mu_\alpha}}{Y_{+}^{(\alpha)}}\right) \hat{\star} K^{y 0}-\sum_{\alpha=1,2} \log \left(1-\frac{e^{i \mu_\alpha}}{Y_{-}^{(\alpha)}}\right) \hat{\star} K^{y 0}\,,
	\end{aligned}
	\vspace{2mm}
\end{equation}
and for auxiliary particles $ \tx{y}^{-} $ and $ \tx{y}^{+} $ the following coupled pair appears
\begin{equation}\label{TBA_4}
\begin{aligned}
	\log Y_{-}^{(\alpha)}&=-\log \left(1+Y_Q\right) \star K_{-}^{Q y}+\log \left(1+\overline{Y}_Q\right) \star K_{+}^{Q y}\\
 &\qquad\qquad\qquad\qquad\qquad\qquad\qquad+ \sum_{\dot{\alpha} = 1}^{N_0} \log \left( 1+Y_0^{(\dot{\alpha})} \right) \check{\star} K^{0 y} \,,
\end{aligned}
\end{equation}
\vspace{2mm}
\begin{equation}\label{TBA_5}
\begin{aligned}
	\log Y_{+}^{(\alpha)} &= - \log \left(1+Y_Q\right) \star K_{+}^{Q y} + \log \left(1+\overline{Y}_Q\right) \star K_{-}^{Q y}\\
 &\qquad\qquad\qquad\qquad\qquad\qquad\qquad- \sum_{\dot{\alpha} = 1}^{N_0} \log \left( 1+Y_0^{(\dot{\alpha})} \right) \check{\star} K^{0 y} \,.
\end{aligned}
\end{equation}
 The mirror energy $ \tilde{\cl{E}}_{Q} $ that depends on the corresponding mirror momentum or $u$-rapidity is given by
\begin{equation}
	\widetilde{\cl{E}}_{Q} = \log \frac{x\left( u - i \frac{Q}{h} \right)}{x\left( u + i \frac{Q}{h} \right)} = 2 \, \tx{arcsinh}\left( \dfrac{\sqrt{ \widetilde{p}^{2} + Q^{2}}}{2 h} \right)\,,
\end{equation}
\begin{equation}
	\widetilde{p}(u,Q) = h \left[ x\left(u - i \frac{Q}{h}\right) - x\left(u + i \frac{Q}{h}\right)\right] + iQ
\end{equation}
whereas the massless counterpart can be obtained in the $ Q \rightarrow 0 $ limit
\begin{equation}
	\widetilde{\cl{E}}_{0} = \log \frac{x\left( u - i 0 \right)}{x\left( u + i 0 \right)} = 2 \, \tx{arcsinh}\left( \dfrac{| \widetilde{p} |}{2 h} \right)\,.
\end{equation}

\section{S-matrix \texorpdfstring{\boldmath$ S^{pq} $}{S**(p,q)} and Kernel \texorpdfstring{\boldmath$ K^{pq} $}{K**(p,q)}}\label{Apx:S_K_Relations}

On {\adso} background different mirror sectors constitute distinct analytic properties \cite{Frolov:2021zyc,Frolov:2021fmj} involved in description of scattering. This is directly reflected in the corresponding kernels $ K^{ab} $ of the TBA system \ref{Apx:TBASystem}. By definition the kernels depend on the associated {\Sx}
\begin{equation}
	K_{ij}(u,v) = \dfrac{1}{2 \pi i} \dfrac{\dd}{\dd u} \log S_{ij}(u,v)
\end{equation}
hence identified by the scattering data. Here we consider the $ u $-parametrisation of an {\Sx} and kernel $ K $, which in the present framework is given by the Zhukovsky relation 
\begin{equation}
	x(u) = \dfrac{1}{2} \left( u - i\sqrt{4-u^{2}} \right)
\end{equation}
where $ x $ is Zhukovsky variable and the relation exhibits long cuts $ -2 > u > 2 $ for $ u \in \bb{R} $. Depending on the scattered sectors, the corresponding cut structure is also reflected in an appropriate convolution bounds, \textit{i.e.}
\begin{equation}
	\star \leftrightarrow \int_{-\infty}^{+\infty} \mathrm{d} u \quad \hat{\star} \leftrightarrow \int_{-2}^{+2} \mathrm{~d} u \quad \check{\star} \leftrightarrow\left(\int_{-\infty}^{-2}+\int_{+2}^{+\infty}\right) \mathrm{d} u 
\end{equation}
The $ \star $-left action that is involved in \ref{Apx:TBASystem} defines for any domain
\begin{equation}
	\rho_{i} \star K_{ij} (v) \equiv \sum_{i} \int \dd u \, \rho_{i}(u) K_{ij}(u,v)
\end{equation}

\noindent The $ S $-matrices that are involved in the TBA and the fused Bethe-Yang system can be grouped into scattering sectors by
\vspace{-3mm}
\paragraph{Massive chiral sector}
Since at the TBA level convolutions that contribute arise analogously for the left and right sectors, we can define them on equal grounds as
\begin{equation}
	\begin{aligned}
		& S_{-}^{Q y}(u, v)=\frac{x(v)-x\left(u-i \frac{Q}{h}\right)}{x(v)-x\left(u+i \frac{Q}{h}\right)} \sqrt{\frac{x\left(u+i \frac{Q}{h}\right)}{x\left(u-i \frac{Q}{h}\right)}} \\[1ex]
		& S_{+}^{Q y}(u, v)=\frac{x\left(u-i \frac{Q}{h}\right)-\frac{1}{x(v)}}{x\left(u+i \frac{Q}{h}\right)-\frac{1}{x(v)}} \sqrt{\frac{x\left(u+i \frac{Q}{h}\right)}{x\left(u-i \frac{Q}{h}\right)}}
	\end{aligned}
\end{equation}
which leads to the $ K_{\mp}^{Qy} $ kernels that arise in the analytically splitted form from the universal kernels $ K $ and $ K_{Q} $. More specifically the unified relation can be compactly given (rapidity dependent)
\begin{equation}
	K_{\mp}^{Q y}(u, v)=\frac{1}{2}\left(K_Q(u-v) \pm K_{Q y}(u, v)\right)
\end{equation}
\begin{equation}
	K_{Q y}(u, v) = K\left(u-\frac{i}{h} Q, v\right)-K\left(u+\frac{i}{h} Q, v\right)
\end{equation}
\begin{equation}
	K_Q(u) = \frac{1}{\pi} \frac{h Q}{Q^2+h^2 u^2} \, , \qquad K(u, v) = \frac{1}{2 \pi i} \frac{\sqrt{4-v^2}}{\sqrt{4-u^2}} \frac{1}{u-v}
\end{equation}
Important that for massive kernels above, we have already implemented the map that is given by (in comparison to \cite{Frolov:2021bwp})
\begin{equation}
	x_{i}^{\pm} = x\left( u_{i} \pm \dfrac{i}{h} \right) \quad u_i=u+\frac{(Q + 1 - 2 i) i}{h}, \quad i = 1, \ldots Q
\end{equation}
where all real particles possess $ u \in \bb{R} $ and analogous for the right sector $ \overline{Q} $. On the other hand, for the bound states $ \{ \widetilde{\cl{E}}, \widetilde{p}_{k},u \} \in \bb{C} $. As it was shown by means of fusion operation \cite{Frolov:2021bwp}, the bound states can be treated as particles of mass gap $ Q \in \bb{N} $.

\vspace{5mm}
\paragraph{Massless sector}
The massless-auxiliary matrices are defined as
\begin{equation}
	S^{0 y}\left(x_k, y_j\right)=e^{+\frac{i}{2} p_k} \frac{\frac{1}{x_k}-y_j}{x_k-y_j}=\frac{1}{S^{0 y}\left(x_k, \frac{1}{y_j}\right)}
\end{equation}
\begin{equation}
	x_{k} = x \left( u_{k} + i \epsilon \right) = x \left( u_{k} - i \epsilon \right)^{-1} \, \qquad \epsilon \rightarrow 0
\end{equation}
where the last identity conventionally based on
\begin{equation}
	e^{i p_{k}} = \dfrac{x^{+}}{x^{-}}
\end{equation}
and the massless-auxiliary sector interchange can occur through braiding-unitarity relation
\begin{equation}
	S^{y 0}(y_{j}, x_{k})=\frac{1}{S^{0 y}(x_{k}, y_{j})}
\end{equation}
All particles in the physical region are real and there is no formation of the bound states in the massless sector. For unified argument, the massless sector can be considered as formal $ Q \rightarrow 0 $ limit.

\vspace{5mm}
\paragraph{Auxiliary sector}
For the reverted sector, \textit{i.e.} the auxiliary $ y $-particles appear first, we obtain
\begin{equation}
	\begin{aligned}
		& S_{-}^{y Q}(u, v)=\frac{x(u)-x\left(v+i \frac{Q}{h}\right)}{x(u)-x\left(v-i \frac{Q}{h}\right)} \sqrt{\frac{x\left(v-i \frac{Q}{h}\right)}{x\left(v+i \frac{Q}{h}\right)}} \\[1ex]
		& S_{+}^{y Q}(u, v)=\frac{\frac{1}{x(u)}-x\left(v-i \frac{Q}{h}\right)}{\frac{1}{x(u)}-x\left(v+i \frac{Q}{h}\right)} \sqrt{\frac{x\left(v+i \frac{Q}{h}\right)}{x\left(v-i \frac{Q}{h}\right)}} 
	\end{aligned}
\end{equation}
Similarly the corresponding kernels follow from universal ones above as
\begin{equation}
	K_{ \pm}^{y Q}(u, v)=\frac{1}{2}\left(K_{y Q}(u, v) \mp K_Q(u-v)\right)
\end{equation}
\begin{equation}
	K_{y Q}(u, v) = K\left(u, v+\frac{i}{h} Q\right)-K\left(u, v-\frac{i}{h} Q\right)
\end{equation}
All auxiliary particles that appear in string hypothesis have $ |y| = 1 $. In this regard, maps differ for negative/positive imaginary parts
\begin{equation}
	\begin{cases}
		y = x(u) \, , \quad &\Im[y]<0\,, \\ 
		y = \frac{1}{x(u)} \, , \quad &\Im[y]>0\,.
	\end{cases}
\end{equation}

\bibliographystyle{JHEP}
\bibliography{References.bib}

\end{document}